\documentclass[letterpaper,twocolumn,10pt]{article}
\usepackage{usenix-2020-09}

\usepackage{tikz}
\usepackage{amsmath}

\usepackage{filecontents}

\usepackage{cite}
\usepackage{amsmath,amssymb,amsfonts}
\usepackage{algorithmic}
\usepackage{graphicx}
\usepackage{textcomp}
\usepackage{xcolor}

\usepackage{color}
\usepackage{hyperref}
\usepackage{graphicx}
\usepackage[inkscapelatex=false]{svg}
\usepackage{multirow} 
\usepackage{subfigure}
\usepackage{makecell} 
\usepackage{booktabs} 
\usepackage{enumitem} 
\usepackage{pifont} 
\usepackage{mdframed} 
\usepackage{colortbl} 

\def\myModel{{\sc Connector}} 
\newcommand{\red}[1]{\textcolor{black}{#1}} 

\usepackage{listings, xcolor}
\definecolor{verylightgray}{rgb}{.97,.97,.97}
\lstdefinelanguage{Solidity}{
  keywords=[1]{anonymous, assembly, assert, balance, break, call, callcode, case, catch, class, constant, continue, constructor, contract, debugger, default, delegatecall, delete, do, else, emit, event, experimental, export, external, false, finally, for, function, gas, if, implements, import, in, indexed, instanceof, interface, internal, is, length, library, log0, log1, log2, log3, log4, memory, modifier, new, payable, pragma, private, protected, public, pure, push, require, return, returns, revert, selfdestruct, solidity, storage, struct, suicide, super, switch, then, this, throw, transfer, true, try, typeof, using, value, view, while, with, addmod, ecrecover, keccak256, mulmod, ripemd160, sha256, sha3}, 
  keywordstyle=[1]\color{blue}\bfseries,
  keywords=[2]{address, bool, byte, bytes, bytes1, bytes2, bytes3, bytes4, bytes5, bytes6, bytes7, bytes8, bytes9, bytes10, bytes11, bytes12, bytes13, bytes14, bytes15, bytes16, bytes17, bytes18, bytes19, bytes20, bytes21, bytes22, bytes23, bytes24, bytes25, bytes26, bytes27, bytes28, bytes29, bytes30, bytes31, bytes32, enum, int, int8, int16, int24, int32, int40, int48, int56, int64, int72, int80, int88, int96, int104, int112, int120, int128, int136, int144, int152, int160, int168, int176, int184, int192, int200, int208, int216, int224, int232, int240, int248, int256, mapping, string, uint, uint8, uint16, uint24, uint32, uint40, uint48, uint56, uint64, uint72, uint80, uint88, uint96, uint104, uint112, uint120, uint128, uint136, uint144, uint152, uint160, uint168, uint176, uint184, uint192, uint200, uint208, uint216, uint224, uint232, uint240, uint248, uint256, var, void, ether, finney, szabo, wei, days, hours, minutes, seconds, weeks, years},  
  keywordstyle=[2]\color{teal}\bfseries,
  keywords=[3]{block, blockhash, coinbase, difficulty, gaslimit, number, timestamp, msg, data, gas, sender, sig, value, now, tx, gasprice, origin},  
  keywordstyle=[3]\color{violet}\bfseries,
  keywords=[4]{[1]},
  keywordstyle=[4]\color{blue}\bfseries,
  identifierstyle=\color{black},
  sensitive=false,
  comment=[l]{//},
  morecomment=[s]{/*}{*/},
  commentstyle=\color{gray}\ttfamily,
  stringstyle=\color{red}\ttfamily,
  morestring=[b]',
  morestring=[b]"
}
\begin{document}

\date{}

\title{{\sc Connector}: Enhancing the Traceability of Decentralized Bridge Applications via Automatic Cross-chain Transaction Association}

\author{
{\rm Dan Lin}\\
Sun Yat-sen University
\and
{\rm Jiajing Wu}\\
Sun Yat-sen University
\and
{\rm Yuxin Su}\\
Sun Yat-sen University
\and
{\rm Ziye Zheng}\\
Sun Yat-sen University
\and
{\rm Yuhong Nan}\\
Sun Yat-sen University
\and
{\rm Qinnan Zhang}\\
Beihang University
\and
{\rm Bowen Song}\\
Ant Group
\and
{\rm Zibin Zheng}\\
Sun Yat-sen University
} 

\maketitle


\begin{abstract}
    Decentralized bridge applications are important software that connects various blockchains and facilitates cross-chain asset transfer in the decentralized finance (DeFi) ecosystem which currently operates in a multi-chain environment.
    Cross-chain transaction association identifies and matches unique transactions executed by bridge DApps, which is important research to enhance the traceability of cross-chain bridge DApps. However, existing methods rely entirely on unobservable internal ledgers or APIs, violating the open and decentralized properties of blockchain. In this paper, we analyze the challenges of this issue and then present \myModel, an automated cross-chain transaction association analysis method based on bridge smart contracts. Specifically, \myModel~first identifies deposit transactions by extracting distinctive and generic features from the transaction traces of bridge contracts. With the accurate deposit transactions, \myModel~mines the execution logs of bridge contracts to achieve withdrawal transaction matching. We conduct real-world experiments on different types of bridges to demonstrate the effectiveness of \myModel. The experiment demonstrates that \myModel~successfully identifies 100\% deposit transactions, associates 95.81\% withdrawal transactions, and surpasses methods for CeFi bridges. 
    Based on the association results, we obtain interesting findings about cross-chain transaction behaviors in DeFi bridges and analyze the tracing abilities of \myModel~to assist the DeFi bridge apps.
\end{abstract}

\section{Introduction}
\label{Introduction}

\red{Decentralized applications (DApps) on the blockchain combine backend smart contracts with frontend user interfaces, allowing users to participate in various decentralized finance (DeFi) activities without the need to trust centralized third-party institutions~\cite{10.1145/3377812.3382178}. Decentralized bridges (referred to as \textit{DeFi bridges}) are a type of DApp that provides cross-chain transaction services. Currently, the blockchain ecosystem consists of multiple chains, with records of 260 public blockchains as of March 2024, according to data from DeFiLlama\footnote{\url{https://defillama.com/chains}}. 
However, data between different blockchain systems are not interoperable, akin to isolated islands. Therefore, DeFi bridges, as applications capable of facilitating asset circulation and information exchange between chains, can promote further development of the multi-chain ecosystem~\cite{10174993}.}

\begin{figure}
\setlength{\abovecaptionskip}{0.2cm}
    \centering
    \vskip -1ex
    \includegraphics[width=0.9\linewidth]{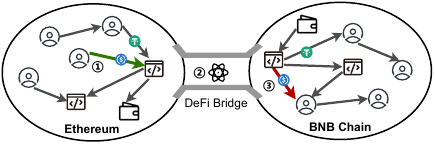}
    \caption{A user making a cross-chain transaction from Ethereum to BNB chain includes three main steps: \ding{172} User locks assets to bridge contract on Ethereum (deposit transaction $tx_{dep}$); \ding{173} The bridge verifies the transaction;
\ding{174} The bridge contract unlocks the corresponding value of the asset to the user on the BNB chain (withdrawal transaction $tx_{wth}$).}
    \label{fig:cross-tx}
\end{figure}

\red{Decentralized bridges do not actually connect the ledgers and assets between different blockchains but rather serve as intermediaries. From the user's view, completing a cross-chain transaction via a DeFi bridge actually consists of two parts: a deposit transaction on the source chain and a withdrawal transaction on the destination chain. Fig.~\ref{fig:cross-tx} shows the process of a user who wants to transfer his UDSC coins from his Ethereum account to his BNB chain account. By the idea of depositing money on the source chain and withdrawing it on the destination chain, bridge DApps enable crypto assets to be used on different blockchains.}

\red{Monitoring the users' behavior and transactions on DeFi bridges is crucial. Specifically, this paper explores the issue of \textbf{\textit{cross-chain transaction association}} (as described in \S\ref{Problem}), which refers to identifying and matching the unique and consistent transactions executed by bridge applications on the source and destination chains to obtain cross-chain transaction pairs. For developers of bridge DApps, firstly, cross-chain transaction association can enhance the traceability of bridge applications~\cite{Traceability2014}, preventing illicit activities such as money laundering by obfuscating and transferring illegal assets through bridge DApps, thereby helping developers reduce compliance risks. According to Elliptic's report~\cite{Elliptic2023}, Ren Bridge has been used for money laundering involving over \$540 million worth of illegal cryptocurrency assets. Furthermore, the lack of effective cross-chain transaction association analysis methods may fail in detecting token leakage attacks of bridge DApps~\cite{DeFiWarder}. For ordinary users, this can improve the experience of bridge users. In conclusion, research on cross-chain transaction association can enhance the reliability and usability of bridge applications themselves.}

\red{Compared to other DApps, DeFi bridges have unique characteristics: 
1) Complexity: DeFi bridges require multi-step interactions of contracts to fulfill multiple business requirements, and these contracts have richer types of functionalities, such as \texttt{Lock}, \texttt{Confirm}, \texttt{Send} etc; 2) Isolation: Multiple contracts of the DeFi bridges need to be deployed separately on mutually isolated source and destination chains for distributed execution. Deposit transactions on the source chain and withdrawal transactions on the destination chain are segregated, as illustrated in Fig.~\ref{fig:cross-tx}. Therefore, cross-chain transaction association faces two main challenges (C):}
\begin{itemize}[leftmargin=*] \itemsep0.6pt
    \item \red{\textbf{C1: Identify complex deposit transactions on the source chain.} The transaction volume of bridge contracts is large, but users' cross-chain deposit transactions are only a part of them and are not explicitly marked on the blockchain. Existing pattern-based transaction identification methods cannot adapt well to complex and diverse cross-chain scenarios.}
    \item \red{\textbf{C2: Match isolated withdrawal transactions on the destination chain.} Due to the isolation characteristics of cross-chain bridge DApps, the process of cross-chain transaction pairs is not fully recorded on any single blockchain ledger, and the relation between source and destination chains is not explicitly recorded on the chain. Therefore, known source chain transactions of cross-chain transactions do not directly get corresponding transactions on destination chains.}
\end{itemize}

In this paper, we propose a novel analysis approach named \myModel~to address \underline{\textbf{c}}r\underline{\textbf{o}}ss-chain tra\underline{\textbf{n}}saction associatio\underline{\textbf{n}} in d\underline{\textbf{ec}}en\underline{\textbf{t}}ralized bridge c\underline{\textbf{o}}nt\underline{\textbf{r}}acts.
We first conduct an empirical study of bridge DApps and summarize the \textit{cross-chain transaction metadata} as the essential elements for cross-chain transaction actions. 
To tackle \textbf{C1}, we combine various aspects of the bridge's input data and execution trace data, extracting semantic and structural features of transaction intent to jointly characterize the intent of cross-chain deposit transactions. 
To tackle \textbf{C2}, we analyze the message-passing mechanism of the bridge at the contract log level, parsing logs for syntax and semantics to extract cross-chain transaction metadata, enabling precise matching in massive transaction data. 
To evaluate the effectiveness of \myModel, we collect the cross-chain transaction pairs from open-source DeFi bridges. We build the first dataset with 24,392 cross-chain transaction pairs from bridges of different cross-chain mechanisms. We conduct experiments on this dataset and successfully identify 100\% deposit transactions and associate 95.81\% withdrawal transactions. Furthermore, based on transaction pairs output by \myModel, we conduct an empirical study in \S\ref{Empirical} to analyze the characteristics of identified deposit transactions, as well as the matched withdrawal transactions to analyze the issues including the user privacy protection, time efficiency, and economic benefits on the cross-chain bridges.
In summary, our main contributions are as follows.
\begin{itemize}[leftmargin=*] \itemsep0.6pt
    \item \textbf{Problem:} To the best of our knowledge, we are the \textit{first} work on cross-chain transaction association tool for decentralized bridge applications. We design \myModel\footnote{Available at {\url{https://github.com/Connector-Tool}}.}~with a more general, multi-cryptocurrency enabled and multi-blockchain-adapted setting. 
    \item \textbf{Evaluation:} \myModel~automatically identify deposit transactions with 100\% accuracy and associate withdrawal transactions with a 95.81\% matching rate. We also apply \myModel~to compare with bridge explorers and \myModel~is validated on 50 unlabelled pairs. 
    \item \textbf{Analysis:} Based on identified deposit transactions and matched withdrawal transactions, we obtain interesting findings through statistical analysis, feature analysis, and graph analysis. Particularly, we find that it is common for ordinary users to use the same address in cross-chain transactions, and DeFi bridges are used in crypto money laundering.
\end{itemize}


\section{Background}
\label{Background}

\subsection{Blockchain and Transactions}
Permissionless blockchains are independent, and their respective ledgers are not interconnected. Ethereum is the first blockchain to support smart contracts, running on the Ethereum Virtual Machine (EVM). Ethereum has two types of accounts: external owned accounts (EOA) - controlled by anyone with the private key, and contract accounts - smart contracts deployed on the network controlled by code.
Smart contracts~\cite{287320,9316905} include code data that can be executed automatically on blockchains. The smart contract code writing in Solidity language~\cite{10.1145/3597926.3598125} can be compiled to generate binary code (bytecode), and also generate the application binary interface (ABI). The ABI records all the functions and events provided by the contract and the parameters that should be passed.
There are two types of transactions: external transactions - initiated by EOA, and internal transactions - initiated by smart contracts. An external transaction can trigger multiple internal transactions.
During the execution of a transaction, the smart contract may trigger defined events into log data. Smart contract events are a special type of structure and are recorded in the EVM's logs. 

\subsection{DeFi Bridges vs. CeFi Bridges}
\begin{figure}[htbp]
\setlength{\abovecaptionskip}{0.2cm}
    \centering
    \includegraphics[width=0.85\linewidth]{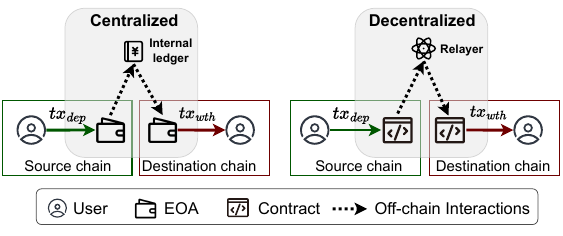}
    \caption{CeFi bridges vs DeFi bridges. } 
    \label{fig:CeFi_DeFi}
\end{figure}

Based on the adopted trust and validation models, cross-chain bridge apps can currently be categorized into centralized bridges (\textit{CeFi bridges}) and decentralized bridges (\textit{DeFi bridges}) 
Fig.~\ref{fig:CeFi_DeFi} illustrates the comparison of the CeFi bridge and DeFi bridges, which have different implementation mechanisms and carriers. 

\noindent\textbf{CeFi bridge:} Most CeFi bridge apps are externally owned accounts (EOA)-based without on-chain code, similar to centralized exchanges (CEX)~\cite{Zhang2022CLTracer}. 
In CeFi bridges, digital assets, and transaction data for cross-chain services are custodied by centralized entities, e.g., ShapeShift~\cite{Yousaf2019Tracing}. 
The centralized entities behind these bridges maintain an internal ledger that records how assets from the source chain are transferred to the destination chain.

\noindent\textbf{DeFi bridge:} DeFi bridge apps are primarily contract-based DApps~\cite{Zhang2022Xscope}, similar to decentralized exchanges.
The on-chain router contract is responsible for interacting with users and various token contracts, providing on-chain functionality, including locking and unlocking users' tokens, and recording token transfer information as specific on-chain events for off-chain relayers to access. Correspondingly, off-chain relayers are responsible for fetching on-chain events from the source chain and coordinating with router contracts on the target chain to complete cross-chain asset transfers. 


\section{Data Source and Statistics}
\label{DateSource}

Current research on cross-chain primarily focuses on blockchain interoperability protocols~\cite{10123097} or generic cross-chain architectures, which are orthogonal to application-level bridge applications we discuss.  
Thus, we try to address a gap in empirical research by examining application-level cross-chain bridges within the market, particularly the trust model (i.e., centralization vs. decentralization) and traceability performance.
We first investigate the websites that currently contain bridge apps, including DefiLlama, Chainspot, and DeFi Rekt Database. 
Chainspot\footnote{\url{https://chainspot.io/}} currently records the highest number of bridges. According to Chainspot's statistics, there are currently \red{120} recorded bridges. We initiate an empirical study to assess the traceability of these bridges, examining various aspects, including their official websites, open-source documentation, verification mechanisms, contract addresses and codes, and transaction retrieval services. 

Regarding the official websites of these cross-chain bridges, \red{17.50\%} bridges do not provide an official website or have discontinued their use. It is important to note that this specific portion of the bridge will not be factored into the subsequent statistics.
Regarding open-source documentation, \red{19.34\%} bridges do not offer white paper or design documentation. 
Regarding the verification mechanisms, \red{4.08\%} bridges are validated by a centralized party, which we considered as CeFi bridges and \red{89.80\%} bridges are validated by multi-validation with trustless code. 
The remaining \red{6.12\%} of bridges cannot validate the validation mechanism due to the lack of documentation on the implementation architecture. 
Among the contract-based DeFi bridges, \red{81.63\%} of bridges disclose their contract addresses and codes. 

\red{Among the available bridges that provided official websites, \red{67.35\%} bridges do not offer cross-chain transaction tracking services or bridge explorers}\footnote{It is worth noting that two potentially confusing scenarios do not qualify as providing cross-chain transaction association services: 1) offering a blockchain explorer instead of a bridge explorer, e.g. Avalanche Bridge; 2) providing an explorer for the latest transactions without supporting history retrieval, e.g. Across.}.  
\red{19.39\% bridges offer cross-chain transaction query services but do not provide detailed cross-chain information. Just \textbf{13.27\% of bridges (13 in total)} offer full cross-chain transaction tracking services, i.e., providing users' deposit and withdrawal transactions.}
There could be several reasons why cross-chain bridges do not offer transaction tracking services: a lack of focus and investment in cross-chain transaction tracking services; limited by the technical complexity and resource constraints; security and privacy considerations.
The comprehensive dataset can be accessed at the open repository.
Among the remaining bridges that do claim to provide tracking services, we observe that they may not consistently return accurate results. For example, in Celer cBridge, a user's cross-chain transaction on Ethereum (deposit transaction hash \href{https://etherscan.io/tx/0x13b1750ee43d05899a350654eaac2a84e5d1001078e40381223368f3dba35b6f}{\textit{0x13b1}}) could not be correctly associated with the withdrawal transaction using the provided service (See \S\ref{Case_study} for more discussion).

\label{Traceability}


\section{Problem Statement and Definition}
\label{Problem}


As described in \S\ref{DateSource}, the traceability of bridge applications refers to the ability to track asset movements and transaction activities after using the bridge apps. Our research focuses on the cross-chain transaction association problem, specifically, identifying and matching the cross-chain transaction pairs of a DeFi bridge between different blockchain platforms.

\noindent{\sc \textbf{Problem} (Cross-chain transaction association):}
\textit{Given a DeFi bridge without centralized ledgers or APIs, we suppose it supports $K$ different and non-interconnected blockchain platforms. The number of the bridge transaction data on the $k$-th chain is $N_k$, and the transaction set in its ledger is denoted as $TX^k=\{tx^k_1, tx^k_2, ...,tx^k_{N_k}\}$. Each transaction is uniquely identified by its transaction hash.}

\noindent$\bullet$ \textit{Identify the deposit transaction $tx^{A}_{i}$ on the source chain $A$;}

\noindent$\bullet$ \textit{Locate corresponding withdrawal transaction $tx^B_{j}$ on destination chain $B$ for each deposit transaction.} 


This process ultimately results in a collection of cross-chain transaction pairs with consistent one-deposit and one-withdrawal behavior\footnote{Inconsistent behavior like attack transactions~\cite{Zhang2022Xscope} and many-to-many cross-chain relationships are out of the scope discussed in this paper.}, denoted as
$\{(tx_i^A,tx_j^B)|tx_i^A \in TX^A,tx_j^B \in TX^B\}.$ For the sake of subsequent description, we first define \textbf{\textit{cross-chain transaction metadata}} (denoted as $\mathbb{M}$) for a cross-chain transaction pair, as shown in Table~\ref{tab:metadata}. 

\begin{table}[h]
\setlength{\abovecaptionskip}{0.2cm}
  \centering
  \caption{Cross-chain transaction metadata ($\mathbb{M}$). Superscript `$s$' indicates source chains, and `$d$' indicates destination chains.} 
  \scalebox{0.63}{
    \begin{tabular}{ll|ll}
    \toprule
    \textbf{Symbol} & \textbf{Description} & \textbf{Symbol} & \textbf{Description} \\
    \midrule
    \rowcolor{gray!10} 
    $\mathbb{M}.txhash^{s}$ & Deposit transaction hash  & $\mathbb{M}.txhash^{d}$ & Withdrawal transaction hash  \\
    $\mathbb{M}.chain^{s}$ & Source chain & $\mathbb{M}.chain^{d}$ & Destination chain \\
    \rowcolor{gray!10} 
    $\mathbb{M}.sender$ & Deposit address & $\mathbb{M}.receiver$ & Withdrawal address \\
    $\mathbb{M}.asset^{s}$ & Deposit asset hash & $\mathbb{M}.asset^{d}$ & Withdrawal asset hash \\
    \rowcolor{gray!10} 
    $\mathbb{M}.amount^{s}$ & Deposit amount & $\mathbb{M}.amount^{d}$ & Withdrawal amount \\
    $\mathbb{M}.timestamp^{s}$ & Deposit timestamp & $\mathbb{M}.timestamp^{d}$ & Withdrawal timestamp \\
    \bottomrule
    \end{tabular}%
  }
  \label{tab:metadata}%
\end{table}%

Cross-chain transaction metadata constitutes the essential data for cross-chain transaction actions in bridge DApps, regardless of the variant mechanisms of bridges, which can be generalized to all types of bridges.
Cross-chain transaction metadata comprises two sets of information from the source and destination chains. The metadata from each chain includes fundamental details of a transaction, such as transaction hash, user address, timestamp, asset hash (i.e., type), and amount.
This metadata is the key to cross-chain transaction identification and matching. Our design idea of \myModel~is to reveal the metadata of cross-chain transactions reliant on bridge smart contracts. The proposed cross-chain metadata are deeply involved in the subsequent deposit transaction identification (See \S\ref{Deposit_Identification}) and withdrawal transaction identification (See \S\ref{Withdrawal_Association}).





\subsection{Existing Work and Limitations}
The comparison of existing cross-chain transaction association methods and ours are shown in Table~\ref{tab:existing_method}.
Yousaf, Kappos, and Meiklejohn (referred to as YKM Heuristics hereafter)~\cite{Yousaf2019Tracing} is the first to study the cross-chain transaction association problem in EOA-based CeFi bridges.
Zhang \textit{et al.}~\cite{Zhang2022CLTracer} devise CLTracer, a framework grounded in address relationships for cross-chain transaction clustering. 
Existing methods cannot be directly applied to DeFi bridges without centralized ledgers or APIs, and still have the following research gaps:
\textit{1) Low accuracy:} The accuracy of the original tracing algorithm proposed in the YKM Heuristics, which relies on bridge APIs to identify deposit transactions on the blockchain, is only 80\%~\cite{Yousaf2019Tracing}, leaving room for improvement.
\textit{2) Non-independent:} Existing methods rely entirely on the internal APIs to match withdrawal transactions. This leads to a lack of user confidence in the traceability of bridges, as internal data can be manipulated by a centralized entity with the potential for tampering and not aligning with the nature of DeFi. This motivates us to design cross-chain transaction association method for DeFi bridges.

\begin{table}[t]
\setlength{\abovecaptionskip}{0.2cm}
\renewcommand{\arraystretch}{1.4}
  \centering
  \caption{Comparison of existing related works.}
  \scalebox{0.5}{
    \begin{tabular}
    {c|cccccc}
    \toprule
    \textbf{\scalebox{1.2}{Solutions}} &
    \multicolumn{1}{p{5em}}{\centering\textbf{\scalebox{1.2}{Supported} \scalebox{1.2}{Types}}} & 
    \multicolumn{1}{p{5em}}{\centering\textbf{\scalebox{1.2}{Bridge} \scalebox{1.2}{Carriers}}} &
    \multicolumn{1}{p{6em}}{\centering\textbf{\scalebox{1.2}{Data} \scalebox{1.2}{Source}}} &
    \multicolumn{1}{p{5em}}{\centering\textbf{\scalebox{1.2}{ERC20} \scalebox{1.2}{Assets?}}} &
    \multicolumn{1}{p{6.5em}}{\centering\textbf{\scalebox{1.2}{Without} \scalebox{1.2}{Bridge APIs?}}} &
    \multicolumn{1}{p{5em}}{\centering\textbf{\scalebox{1.2}{Support} \scalebox{1.2}{History?}}} \\
    \midrule
    \rowcolor{gray!10} 
    \scalebox{1.3}{YKM ~\cite{Yousaf2019Tracing}} & \multicolumn{1}{p{6em}}{\centering \scalebox{1.2}{CeFi Bridge} \scalebox{1.2}{Shapeshift}} & \scalebox{1.2}{EOA}  & \multicolumn{1}{p{6em}}{\centering \scalebox{1.2}{Synchronized} \scalebox{1.2}{full ledgers}}  & \scalebox{2}{×}     & \scalebox{2}{×}     & \scalebox{2}{×} \\
        
    \scalebox{1.3}{CLTracer~\cite{Zhang2022CLTracer}} & \multicolumn{1}{p{6em}}{\centering \scalebox{1.2}{CeFi Bridge/} \scalebox{1.2}{CEX}} & \scalebox{1.2}{EOA}  & \multicolumn{1}{p{6em}}{\centering \scalebox{1.2}{Synchronized} \scalebox{1.2}{full ledgers}}  & \scalebox{2}{×}     & \scalebox{2}{×}     & \scalebox{1.5}{\checkmark} \\
    \rowcolor{gray!10} 
    \scalebox{1.3}{Ours}  & \scalebox{1.2}{DeFi Bridge} & \scalebox{1.2}{Contract}  & \multicolumn{1}{p{6em}}{\centering \scalebox{1.2}{Crawling} \scalebox{1.2}{limited data}}  & \scalebox{1.5}{\checkmark}     & \scalebox{1.5}{\checkmark}     & \scalebox{1.5}{\checkmark} \\
    \bottomrule
    \end{tabular}%
}
  \vskip -1ex 
  \label{tab:existing_method}%
\end{table}%

%



\section{Methodology of \myModel}
\label{Methodology}

\begin{figure}[h]
\setlength{\abovecaptionskip}{0.2cm}
    \centering
    \includegraphics[width=1\linewidth]{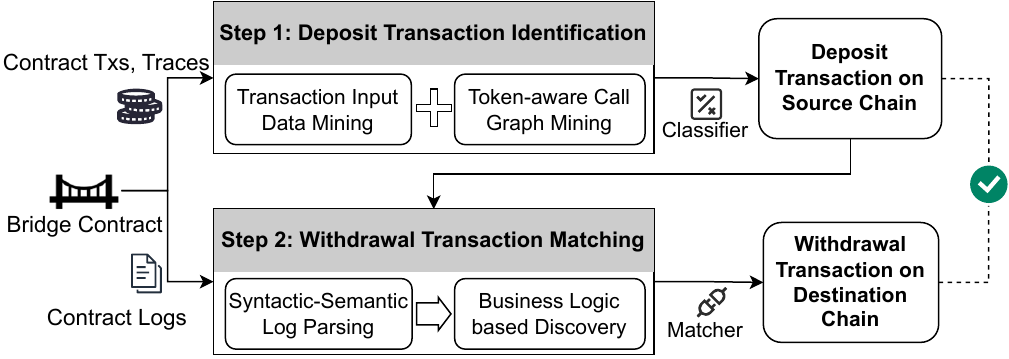}
    \caption{Overview of \myModel.}
    \label{fig:XLinker_overview}
\end{figure}

In this section, we design \myModel~to address cross-chain transaction association for DeFi bridges. As shown in Fig.~\ref{fig:XLinker_overview}, \myModel~consists of two main steps. \myModel~takes the transaction data, traces, and logs from the cross-chain bridge contract as input and outputs the bridge's transaction pairs.
The first step (\textbf{S1: Deposit Transaction Identification}) extracts effective features by mining input data and token-aware call relationships of contract transactions for deposit behavior identification. The second step (\textbf{S2: Withdrawal Transaction Matching}) parses the contract log from both syntactic and semantic view, and match the withdrawal transaction via business logic. 

\subsection{Deposit Transaction Identification}
\label{Deposit_Identification}

In bridge applications, the first step for users in conducting cross-chain transactions typically involves initiating a deposit transaction on the source chain. Only by knowing the deposit transactions on the source chain can we further match the corresponding withdrawal transactions. The purpose is to intelligently and precisely distinguish transactions that belong to cross-chain deposit activities, ensuring the effectiveness and efficiency of subsequent cross-chain transaction correlation.
Existing methods identify transactions by manually pre-collecting function names keywords and conduct corresponding method ID database for lookup~\cite{DAppHunter2023,9152785}.
However, the major challenge with this approach lies in its lack of generality, which can lead to false positives. 
1) It struggles to handle newly added functions during bridge contract upgrades;
2) It faces difficulties in adapting to all diverse bridge contracts.

Deposit transactions in DeFi bridges are implemented through complex and diverse smart contracts. To tackle these challenges, we observe that deposit transactions exhibit two distinct features: one is the \textit{functional} features of transaction input data, and the other is the \textit{structural} features of the token-aware call graph. Those two features will be fed to train a classifier to identify deposit transactions.



\subsubsection{Transaction Input Data Mining}
\label{word2vec}
The most straightforward way to understand a user's deposit transactions is to start by analyzing the interaction transactions between the user and the bridge contract. 
We first investigate the deposit functions and parameters of different bridges with three bridges as an example, as shown in Fig.~\ref{fig:deposit_functions}. 
Different deposit functions for different bridges essentially inform the bridge contract through a transaction to transfer to other blockchains by a certain amount and asset type. 
To address this problem, we intend to apply natural language processing (NLP) to characterize the functional characteristics of bridge transactions. Using NLP tools, the semantic similarities of function definition of bridge contracts can be extracted as a distinguishable feature. Specifically, \myModel~extracts functional features by decoding transaction data and embedding the normalized text sequence.

\begin{figure}[tbp]
\setlength{\abovecaptionskip}{0.2cm}
\begin{lstlisting}[numbers=none]
//Poly Network
function lock(address fromAsset, uint64 toChainId, bytes toAddress, uint256 amount, uint256 fee, uint256 id);
//Celer cBridge
function send(address _receiver, address _token, uint256 _amount, uint64 _dstChainId, uint64 _nonce, uint32 _maxSlippage);
function sendNative(address _receiver, uint256 _amount, uint64 _dstChainId, uint64 _nonce, uint32 _maxSlippage);
//Multichain
function anySwapOutNative(address token, address to, uint256 toChainID);

\end{lstlisting}
\caption{Example deposit functions defined in Poly Network, Celer cBridge, and Multichain. } 
\label{fig:deposit_functions}
\vskip -3ex
\end{figure}

\noindent\textbf{Input data decoding.} The transaction input data are encoded in a hexadecimal string with semantic gaps for human reading and NLP tools. 
The input data specifies the function to be called (i.e., the function selector) in the first four bytes, followed by a list of encoded parameters (starting with the 5th byte).
To decode the input data, \myModel~utilizes the open-source web3 package \texttt{decode\_function\_input} method based on the templates provided in the ABI (introduced in \S\ref{Background}). 
The ABI matches the called function and parameters in the transaction data with human-readable interface definitions. Formally, we get 
$f(<p_i:r_i>),$
where $f$ is the function name, $<\cdot>$ is the list form, $p_i$ is the $i$-th parameter and $r_i$ is the results of value.

\noindent\textbf{Vocabulary embedding.}
Next, we convert decoded input data into vectors to text sequences for extracting transaction semantics. We find the parameters in deposit transactions of different bridges have irrelevant variables other than cross-chain metadata, which may affect the effectiveness of semantic extraction and reduce the performance of our deposit identification method. This inspires us to normalize the decoded data of transactions by mapping parameters other than cross-chain metadata to constant variables.
Our renaming rules are as follows: 
1) Remain the original function name $f$ and ignore all values $v_i$ of the parameters;
2) If the parameter name is similar to the cross-chain metadata element (introduced in \S\ref{Problem}), rename it as the corresponding element. e.g., rename ``fromAsset'' as $asset^s$ (See \href{https://github.com/Connector-Tool}{GitHub} for more renaming rules). All other extraneous parameter names are normalized to \texttt{var\_0}, \texttt{var\_1}, etc., in the order of occurrence. 
\vspace{-1ex}
\begin{equation}
\begin{aligned}
\footnotesize 
    & Normalized(f(<p_i:r_i>))~=~f, <p_i > \\
    & p_i=\left\{ 
    \begin{array}{lc}
        element, & if~p_i \approx element, element \in \mathbb{M} \\
        \texttt{var}, & otherwise
    \end{array} 
    \right.
\end{aligned}
\end{equation}

Finally, we train a word embedding model word2vec~\cite{Mikolov2013Efficient} to get functional features. 
For the normalized transaction set $S=\{s\}, s=f~or~p_i$, we get the word embeddings by $W=Word2Vec(S)$. The word2vec in \myModel~is trained with dimension of word vector \texttt{vector\_size=6}. 
By converting each normalized text to word embedding, \myModel~extracts the functional feature $v_{txhash}^{func}$ of each transaction with a fixed length of 48 (features with insufficient length are filled with zeros). 
To this end, the functional features with semantics are extracted by \myModel~for deposit behavior identification.

\subsubsection{Token-aware Call Graph Mining}


In DeFi bridge apps, different components are deployed on different blockchains with their scope of work and communicate via function calling. The on-chain router smart contracts are responsible for user interactions and various token contracts~\cite{Zhang2022Xscope}.
In this part, we explore the call relationships between users and contracts, which may give rise to distinctive structural features for deposit behavior identification. 

To accomplish users' deposit behavior, the bridge contracts need to trigger several internal function calls to transfer tokens (i.e., assets) between contracts afterwards.
Fig.~\ref{fig:triggered_contracts} shows the function call relationship of an example deposit transaction \href{https://etherscan.io/tx/0xa7ef44e44db882378e1339392f200850dfbf0d817ca271ebf96400df9ef2209e}{\textit{0xa7ef}}.
in Poly Network bridge. 
1) The user first conducts an external transaction to call function \texttt{lock()} of contract \texttt{PolyWrapper}; 
2) Within contract \texttt{PolyWrapper}, the token transfer function \texttt{safeTransferFrom()} is called internally;  
3) After that, another contract \texttt{LockProxy} of the Poly Network bridge is triggered at the entry of the function \texttt{lock()}, which subsequently calls token transfer function \texttt{\_transferToContract()} to further transfer token to the contract \texttt{LockProxy}. 
These call relationships in the bridge transaction can be modeled and mined to find more general features for deposit transaction identification.


\begin{figure}[t]
\setlength{\abovecaptionskip}{0.2cm}
\begin{lstlisting}[numbers=none]
//PolyWrapper.sol
function lock(address fromAsset, uint64 toChainId, bytes memory toAddress, uint amount, uint fee, uint id) external payable nonReentrant whenNotPaused { //[1] external transactions
    ...//requirements
    IERC20(fromAsset).safeTransferFrom(msg.sender, address(this), amount); //[2] ERC20 token transfer
    amount = _checkoutFee(fromAsset, amount, fee);
    address lockProxy = _getSupportLockProxy(fromAsset, toChainId);
    ...//approves
    require(lockProxy.lock(fromAsset, toChainId, toAddress, amount), "lock erc20 fail"); // Call the lock function of contract lockProxy
    ...}
//LockProxy.sol
function lock(address fromAssetHash, uint64 toChainId, bytes memory toAddress, uint256 amount) public payable returns (bool) {
    ...//requirements
    require(_transferToContract(fromAssetHash, amount), "transfer asset from fromAddress to lock_proxy contract  failed!"); //[3] ERC20 token transfer
    ...//cross-chain manager called }

\end{lstlisting}
\caption{The simplified implementation of triggered contracts and functions corresponding to the example transaction \textit{0xa7ef}.}
\label{fig:triggered_contracts}
\end{figure}

    
\begin{figure}[t]
\setlength{\abovecaptionskip}{0.2cm}
    \centering
    \includegraphics[width=\linewidth]{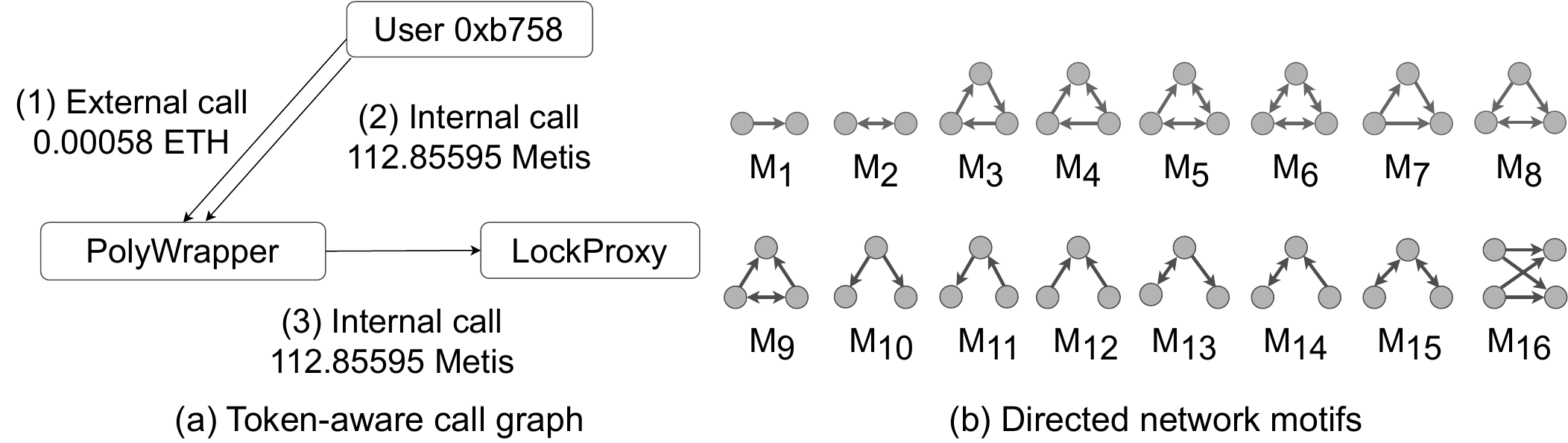}
    \caption{(a) The user \textit{0xb758} transferred 0.00058 ETH to call the contract \texttt{PolyWrapper} in the external transaction; Then the token transfer function is called while 112.8556 Metis Token is transferred to contract \texttt{PolyWrapper}; Finally, those tokens are further transferred to contract \texttt{LockProxy} via its token transfer function. (b) M$_{1}$ and M$_{2}$ are all connected two-vertice motifs; M$_{3}$ – M$_{15}$ are all 13 connected three-vertice motifs; M$_{16}$ is the four-vertice bi-fan motif.}
    \label{fig:callgraph+motifs}
\end{figure}


\noindent\textbf{Modeling of token-aware call graph.} We model the call relationships among each transaction as a call graph. 
In particular, we focus on token-aware call relationships, since deposit behaviors of cross-chain transactions inevitably involve asset transfers. 
Token-aware call relationship is denoted as $(caller,~callee,~call~type,~tokens)$. 
We take the Ethereum deposit transaction \textit{0xa7ef} in Poly Network as an example, and its token-aware call graph is visualized in Fig.~\ref{fig:callgraph+motifs}(a).
The token-aware call graph of this transaction contains three adresses $\{\href{https://etherscan.io/address/0xb758b6576221a7504a7211307092c23d3ee191c9}{0xb757},~PolyWrapper,~LockProxy\}$ and three call relationships 
$\{(0xb757$, $PolyWrapper,$ $external~call,$ $0.00058~Ether),$ 
$(0xb757,$ $PolyWrapper,$ $internal~call,$ $112.86~Metis),$ 
$(PolyWrapper,$ $LockProxy,$ $internal~call,$ $112.86~Metis)\}$. 





%

\noindent\textbf{Motif of token-aware call graph.} With a constructed token-aware call graph, we extract the structural features to characterize call relationship patterns for each transaction.
Network motifs are high-order network organizations, which is an effective complex network tool that can reflect and extract hidden information of the network's structure~\cite{Benson2016Higher}. 
Fig.~\ref{fig:callgraph+motifs}(b) shows 16 network motifs consisting of two and three vertices and a special four-vertice motif. 
These motif patterns have been extensively explored in previous research and have been demonstrated to unveil the distinctive attributes of diverse networks. 
Inspired by MoTs~\cite{Wu2023MoTs}, a generic transaction semantic representation, we apply the concept of network motifs to characterize the transaction patterns of the token-aware call relationship patterns. 
For each transaction ${txhash}$, a 16-dimensional feature vector $v_{txhash}^{struc}$ is extracted. The $i$-th element of $v_{txhash}^{struc}$ means the frequency of the $i$-th network motifs (shown in Fig.~\ref{fig:callgraph+motifs}(b)) in a constructed call graph. We calculate the directed motif M$_{1}$-M$_{16}$ by by subgraph matrix computation~\cite{Benson2016Higher,Wu2023MoTs}. To this end, the structural features with call relationships are extracted by \myModel~for deposit behavior identification.




\subsubsection{Summary for S1}


Two distinct features, the functional feature of transaction input data and the structural features of the token-aware call graph, are concatenated 
to get a more precise and general characterization. 
Note that input data rather than logs are decoded for deposit identification because the former requires less time and storage.
With the extracted features, machine-learning classification tools are employed to identify the deposit behaviors of bridge apps.
Thus, the accurate deposit transactions $\mathbb{M}.txhash^{s}$ of metadata are identified. Also, $\mathbb{M}.sender$, $\mathbb{M}.timestamp^{s}$, $\mathbb{M}.asset^{s}$ of metadata can be extracted from the deposit transaction data and trace.

\subsection{Withdrawal Transaction Matching}

\label{Withdrawal_Association}


As mentioned in \S\ref{Introduction}, inter-blockchain communication within DeFi bridges is executed through distributed smart contracts rather than centralized institutions. Consequently, semantic gaps exist when association transactions within DeFi bridges. To address this challenge, we analyze the execution logs of bridge contracts to identify withdrawal transactions. Specifically, we employ syntax-semantic parsing on deposit transaction logs to extract cross-chain metadata. Subsequently, we identify the target withdrawal transactions based on the business logic of the bridges.




\subsubsection{Syntactic-Semantic Log Parsing}


The contract log serves as communication proof between the source chain and the destination chain, encompassing vital information pertaining to cross-chain metadata. In a DeFi bridge, the general flow of the message-passing mechanism is as follows: when the DeFi bridge receives a cross-chain request, the router contract on the source chain calls the token contract to lock the user's tokens. During this call, the token contract triggers a lock action and passes the user's tokens to the router contract. Subsequently, the router contract generates a deposit event $E$ containing detailed information about the locking behavior, including information about the asset type and amount, as \textit{proof} of the locked asset. An example of a deposit event of Poly Network bridge is: 
\begin{lstlisting}[numbers=none]
event LockEvent(address fromAssetHash, address fromAddress, uint64 toChainId, bytes toAssetHash, bytes toAddress, uint256 amount);
\end{lstlisting}
Next, the off-chain relayers will parse the event and confirm the locking operation on the source chain, then authorize the unlocking operation and send the transaction to the router contract on the destination chain. The router contract will then invoke the target token contract and unlock the asset to the user's address on the destination chain.



\begin{table*}[bt]
\setlength{\abovecaptionskip}{0.2cm}
  \centering
  \caption{The mapping of transaction logs and cross-chain transaction metadata in the semantic parsing.}
  \scalebox{0.62}{
    \begin{tabular}{c|p{64em}|p{5em}}
    \toprule
    
\textbf{Bridge} & \textit{\textbf{Conditions (event + parameter mapping)}} & \multicolumn{1}{l}{\textit{\textbf{\makecell[c]{ Metadata ($\mathbb{M}$)}}}} \\
    \midrule
    \multirow{3}[2]{*}{Celer cBridge} & (Exist \texttt{Send} event $E$ and $E.sender$ = $\mathbb{M}.sender$ and $E.receiver$ = $\mathbb{M}.receiver$ and $E.token$ = $\mathbb{M}.asset^s$ and $\texttt{Decimal(}E.amount\texttt{)}$ = $\mathbb{M}.amount^s$ and $\texttt{ID2Chain(}E.dstChainId\texttt{)}$ = $\mathbb{M}.chain^d$) & \multicolumn{1}{p{5em}}{\multirow{6}[6]{*}{\makecell[l]{$\mathbb{M}.sender$,\\ $\mathbb{M}.receiver$,\\ $\mathbb{M}.asset^s$,\\ $\mathbb{M}.amount^s$,\\ $\mathbb{M}.chain^d$,\\ $\mathbb{M}.asset^d$ $\ast$ }}} \\  
    \multicolumn{1}{c|}{} & or (Exist \texttt{Deposited} event $E$ and $E.depositor$ = $\mathbb{M}.sender$ and $E.mintAccount$ = $\mathbb{M}.receiver$ and $E.token$ = $\mathbb{M}.asset^s$ and $\texttt{Decimal(}E.amount\texttt{)}$ = $\mathbb{M}.amount^s$ and $\texttt{ID2Chain(}E.mintChainId\texttt{)}$ = $\mathbb{M}.chain^d$) &  \\
    \multicolumn{1}{c|}{} & or (Exist \texttt{LogNewTranserOut} event $E$ and $E.sender$ = $\mathbb{M}.sender$ and $E.dstAddress$ = $\mathbb{M}.receiver$ and $E.token$ = $\mathbb{M}.asset^s$ and $\texttt{Decimal(}E.amount\texttt{)}$ = $\mathbb{M}.amount^s$ and $\texttt{ID2Chain(}E.dstChainId\texttt{)}$ = $\mathbb{M}.chain^d$) &  \\
    \cmidrule{1-2}    
    
    Multichain & Exist \texttt{LogAnySwapOut} event $E$ and $E.from$ = $\mathbb{M}.sender$ and $E.to$ = $\mathbb{M}.receiver$ and $E.token$ = $\mathbb{M}.asset^s$ and $\texttt{Decimal(}E.amount\texttt{)}$ = $\mathbb{M}.amount^s$ and $\texttt{ID2Chain(}E.toChainID\texttt{)}$ = $\mathbb{M}.chain^d$ &  \\
    \cmidrule{1-2}    
    
    Poly Network & Exist \texttt{LockEvent} event $E$ and $E.fromAddress$ = $\mathbb{M}.sender$ and $E.toAddress$ = $\mathbb{M}.receiver$ and $E.fromAssetHash$ = $\mathbb{M}.asset^s$ and $E.toAssetHash$ = $\mathbb{M}.asset^d$ and $E.amount$ = $\mathbb{M}.amount^s$ and $\texttt{ID2Chain(}E.toChainId\texttt{)}$ = $\mathbb{M}.chain^d$ &  \\
    \bottomrule
    \multicolumn{3}{l}{\small $\ast$ denotes the metadata element is optional in deposit event logs.}\\
    
    \end{tabular}%
    }
  \label{tab:pattern_mapping}%
\end{table*}%


\noindent\textbf{Syntactic parsing.} To bridge the semantic gap, \myModel~first conducts log syntax parsing to recover the encoded transaction logs. 
In particular, \myModel~utilizes the \texttt{eth\_getTransactionReceipt} function of the source chain to retrieve the receipt of a deposit transaction. 
Then, \myModel~analyzes the logs to detect the events emitted during the execution of deposit transactions. The structured log list is parsed by \myModel~from transaction receipts based on ABIs, which contains all the event interface definitions (i.e., names and parameters).
However, the structured log list itself is undirect to semantic cross-chain metadata information, thus requiring semantic parsing.

\noindent\textbf{Semantic parsing.} 
Semantic parsing aims to map the structured log lists of deposit transactions to cross-chain metadata. 
According to whether the logs meet the conditions of event name and parameters shown in Table~\ref{tab:pattern_mapping}, \myModel~maps the log lists into the semantic cross-chain transaction metadata. 
\texttt{Decimal()} maps the value to the token amount according to different token decimals. 
DeFi bridge apps usually support various tokens, thus it is necessary to unify the value of different tokens for further matching. 
\texttt{ID2Chain()} maps the blockchain ID of different bridges to the blockchain names according to their open document. This is typically seen in bridges that support cross-chaining between multiple blockchains.
Take the deposit transaction \textit{0xa7ef} in Poly Network as an example. 
Based on the log list, \myModel~locates the event mentioned in the condition and maps the parsed parameters to the metadata. Metadata 
($\mathbb{M}$) includes the receiver address on the deposit transaction $\mathbb{M}.receiver$, the sent asset hash $\mathbb{M}.asset^s$, the sent amount $\mathbb{M}.amount^s$, the destination chain $\mathbb{M}.chain^d$, and the timestamp of deposit transaction $\mathbb{M}.timestamp^s$. 
Combining the metadata given by S1, the cross-chain metadata of deposit transaction \textit{0xa7ef} for further matching is:
\vspace{-1ex}
$$ \footnotesize{ \mathbb{M}=\left\{
\begin{array}{rcl}
sender & = & \href{https://etherscan.io/address/0xb758B6576221a7504A7211307092C23D3eE191c9}{\textit{0xb758}} \\
timestamp^s & = & 1680341603 \\
asset^s & = & \href{https://etherscan.io/address/0x9E32b13ce7f2E80A01932B42553652E053D6ed8e}{\textit{0x9E32}}(token) \\
amount^s & = & 112.855947137612614726 \\
chain^d & = & Binance~Smart~Chain \\
asset^d & = & \href{https://bscscan.com/address/0x9E32b13ce7f2E80A01932B42553652E053D6ed8e}{\textit{0x9E32}}(token) \\
receiver & = & \href{https://bscscan.com/address/0xb758b6576221a7504a7211307092c23d3ee191c9}{\textit{0xb758}}
\end{array}
\right.
}
$$
\vspace{-1.2ex}

In this way, \myModel~provides cross-chain transaction metadata containing information of the destination chain by syntactic-semantic log parsing. 


\subsubsection{Business Logic based Target Discovery} 
\label{TargetDiscovery}
In this part, we intend to discover the target withdrawal transaction based on the business logic of bridges. We first implement an efficient search space constructor to obtain candidate transactions, and then design a heuristic matcher among those candidate transactions to discover the withdrawal transaction.



\noindent\textbf{Search space constructor.} The search space is constructed based on the assumption that the cross-chain transactions are usually finished soon in a short time interval, which is the business logic of bridges.
First, \myModel~uses the most recent block number corresponding to $\mathbb{M}.timestamp^{s}$ and $\mathbb{M}.timestamp^{s} + \Delta$ ($\Delta$ denotes the maximum time interval) as the starting block number and the end block number of the search, respectively. 
Then, the search space is constructed by \myModel~crawling all transactions related to $\mathbb{M}.receiver$ that occur between this interval on $\mathbb{M}.chain^d$. In this process, different types of bridges can choose different $\Delta$, and a detailed discussion of $\Delta$ is given in \S\ref{RQ2}.



\noindent\textbf{Heuristic matcher.}
With the obtained search space, \myModel~then discovers the target withdrawal transaction within the search space with the following rules:

\textbf{{\sc Rule 1} (Asset type matching).}
    Based on $\mathbb {M}.asset^{s}$and $\mathbb{M}.asset^{d}$, \myModel~filters candidates with matching asset types. That is, if and only if $\mathbb{M}.asset^{s}$ and $\mathbb{M}.asset^{d}$ are the same, the transaction type matching is satisfied.  

\textbf{{\sc Rule 2} (Transaction time matching).}
    Based on the business logic of cross-chain transactions claimed by bridge apps, withdrawal transactions usually finish within 30 minutes (or less) after the deposit transaction is confirmed. 
    Based on $\mathbb{M}.timestamp^{s}$ and $\mathbb{M}.timestamp^{d}$, \myModel~filters candidates whose time constraint $\tau$ is smaller than the threshold.  
    Initially, \myModel~takes $\tau=30$ as the time-matching threshold and adds 10\% when there are no candidates. Continuing in this manner, the process will iterate until there are transactions that meet the condition. 
    
\textbf{{\sc Rule 3} (Transaction amount matching).}
    Based on the business logic of cross-chain transactions, the cross-chain transaction fee usually accounts for about 0\%-3\% of the deposit transaction amount.
    According to $\mathbb{M}.amount^{s} $ and $\mathbb{M}.amount^{d}$, \myModel~calculates the proportion of transaction fee $\delta$ through $\delta=\frac{\mathbb{M}.amount^{s}-\mathbb{M}.amount^{d}}{\mathbb{M}.amount^{s}},$
    and select candidates whose $\delta \in$ [0\%, 3\%]. Similar to {\sc Rule 2}, due to the special situation where $\delta$ may be greater than 3\%, \myModel~uses 3\% as the lower limit of $\delta$ and 100\% as the upper limit of $\delta$ to gradually relax the amount constraint. 


In most cases, after executing the above matching rules in sequence, the number of candidate withdrawal transactions obtained is 1. 
If there is no candidate, the search space can be further expanded by increasing the $\Delta$. If there is more than 1 candidate, the constraint effect can be enhanced by lowering the $\tau$ of {\sc Rule 2}. Finally, \myModel~outputs a matching pair of deposit transactions and withdrawal transactions.

\subsubsection{Summary for S2}
After syntactic-semantic log parsing, \myModel~reveals the cross-chain metadata information benefit from the message-passing mechanism of DeFi bridges. 
Note that logs instead of input data are parsed for withdrawal transaction matching because the former contains richer essential metadata.
With the metadata information, the search space of the destination chain can be purposefully constructed and discover the target withdrawal transactions with heuristic rules.

\section{Evaluation Results}

\label{Dataset}

Existing mainstream mechanisms for cross-chain bridge include: \textit{Hash Time Lock Contract (HTLC)}, \textit{Notary Mechanism}, and \textit{Relay/Relay chain}~\cite{10123097}. For these mechanisms, we select representative bridges based on the following criteria and gather contract and transaction data as our primary sources for experimentation.
1) the top 12 bridges with the highest liquidity in Q1 2023~\cite{Liquidity}; 2) available bridge explorer service for constructing ground truth dataset (as investigated in \S\ref{DateSource}); 3) EVM-compatible blockchain support; 4) experienced security incidents. Top 3 bridges that satisfy those conditions: Celer cBridge for HTLC, Multichain for Notary Mechanism, and Poly Network for Relay/Relay chain. 
Note that \myModel~discussed in this study is applicable to other contract-based, non-privacy preserving, EVM compatible DeFi bridges.
We then collect contracts and transactions of those bridges. Firstly, our experiments are based on Ethereum as the source chain, since Ethereum is the largest permissionless blockchain platform supporting smart contracts. Ethereum contract addresses are obtained from the bridges' official websites. Then we collect contract transactions from Sept. 2020 to Apr. 2023. Based on collected transactions, we analyze and label the categories, which are divided into deposit and non-deposit transactions. The data statistics are shown in Table~\ref{tab:deposit_dataset}.
Next, we construct a large-scale dataset to obtain the ground truth of cross-chain transaction association.
Based on the deposit transactions, we query the bridge's official explorer to obtain the corresponding withdrawal transactions. We conduct a statistical analysis of the blockchains that had the highest transaction volume in common. The results show that the Top 3 blockchains are Ethereum (ETH), Binance Smart Chain (BSC), and Polygon (MATIC). Thus, we will evaluate the cross-chain \myModel~from Ethereum to Binance Smart Chain and Polygon. 

\begin{table}[t]
\setlength{\abovecaptionskip}{0.2cm}
\renewcommand{\arraystretch}{1.3}
  \centering
  \vskip -2ex
  \caption{Data (Ethereum) for deposit behavior identification.}
  \scalebox{0.6}{
    \begin{tabular}{c|cccc}
    \toprule
    
    \textbf{\scalebox{1.1}{Bridges}} & \textbf{\scalebox{1.1}{\# Contracts}} & \textbf{\scalebox{1.1}{\# Deposit Txs}} & \textbf{\scalebox{1.1}{\# Non-deposit Txs}} & \textbf{\scalebox{1.1}{Sum}} \\
    \midrule
    \rowcolor{gray!10} 
    \scalebox{1.1}{Celer cBridge} & \scalebox{1.1}{4} & \scalebox{1.1}{7,171} & \scalebox{1.1}{12,823} & \scalebox{1.1}{19,994}  \\
    \scalebox{1.1}{Multichain}  & \scalebox{1.1}{10} & \scalebox{1.1}{15,483}  & \scalebox{1.1}{14,512} & \scalebox{1.1}{29,995} \\
    \rowcolor{gray!10} 
    \scalebox{1.1}{Poly Network} & \scalebox{1.1}{3} & \scalebox{1.1}{19,242} & \scalebox{1.1}{148} & \scalebox{1.1}{19,390} \\

    \bottomrule
    \end{tabular}%
   }
  \vskip -3ex
  \label{tab:deposit_dataset}%
\end{table}%

\subsection{Results of Deposit Identification (S1)}
\label{RQ1}
In this part, we first evaluate the effectiveness of \myModel~in identifying deposit transactions. 
To explore generalization, we partition the data from the three bridge contracts into training and testing sets. The testing bridge is \textit{not} available during the training process, and the training set \textit{only} contains data from the remaining two bridges. 
For instance, we train the transaction data of Celer cBridge and Multichain  (19,994+29,995=49,989) to test the transaction data of Poly Network (19,390). The proportion of training and testing sets is 72:28.
To validate the effectiveness of the extracted features, we employ several classical classifiers in the downstream process and quantify the identification performance using accuracy. These classifiers include Logistic Regression (LR), SVC, Decision Tree (DT), Random Forest (RF), AdaBoost, and Gaussian NB (GN), implemented via scikit-learn~\cite{scikit-learn}.
To assess the importance of features, we separately implement: 1) using only functional features; 2) using only structural features; and 3) using both structural features and functional features (i.e. default \myModel).

Table~\ref{tab:M1_results} demonstrates the experimental results. 
\myModel~successfully identifies almost 100\% deposit behavior on average using both structural and functional features. 
With both features, the accuracy variance of our approach among different classifiers is close to 0. 
\red{\myModel~ultimately adopts the AdaBoost classifier~\cite{freund1997decision}, which performs best on both structural and functional features.}
Using only structural features, it can only identify 74.43\% deposit behavior on average, with stable performance.
Using only functional features, it can identify deposit behavior more accurately but is unstable with higher variance.
We will explain the possible reasons below.
Furthermore, to demonstrate the versatility of \myModel~across a wide range of DeFi bridges, we conduct an additional experiment by collecting 10 additional contracts and transactions for different Bridges.
We randomly collect 10 deposit and 10 non-deposit transactions for each additional bridge. 
We use the original data from the three bridges as the training set and Table~\ref{tab:M1_results} demonstrates the results.
Based on the structural features alone, it is stable and can still identify 77\% of deposit transactions.
The functional-based method can only identify 44\% of the deposit behavior.
Our \myModel~combining both features can still maintain the accuracy rate of 87.5\% to 100\% in a wider range of bridges. 
The results indicate that \myModel~can accurately and stably identify deposit behavior. 
\myModel~combining both features excels by combining explicit semantic features -- functional features from input data (ensuring accuracy in similar scenarios), with implicit semantic features -- structural features from call relationships (ensuring generality in dissimilar cases), which enhances overall identification performance.

\begin{table}[t]
\setlength{\abovecaptionskip}{0.2cm}
  \centering
  \caption{Results in deposit transaction identification.}
  \scalebox{0.63}{
    \begin{tabular}{c|c|ccccc|c}
    \toprule
    \multirow{2}[2]{*}{\textbf{Testing}} & \multicolumn{1}{c|}{\multirow{2}[2]{*}{\textbf{Features}}} 
    & \multicolumn{5}{c|}{\textbf{Evaluating Classifiers (Accuracy \%)}} &   \\
          &       & \textbf{LR} & \textbf{SVC} & \textbf{RF} & \textbf{AdaBoost} & \textbf{GN} & \textbf{Avg.}    \\
    \midrule
    
    \multirow{3}[2]{*}{Poly Network$\dagger$} 
          
          & Struc. & 72.97 & 75.16 & 75.16 & 75.16 & 72.97 & 74.43 \\
          & Func. & 100   & 100   & 50.11  & 100   & 48.94 & 83.18 \\
          & Struc. + Func. & 100   & 100   & 99.98 & 100   & 100   & 99.97 \\
    \midrule
    \multirow{3}[2]{*}{Multichain$\dagger$} 
          & Struc. & 72.51 & 74.72 & 74.72 & 74.72 & 72.51 & 73.98 \\
          & Func. & 100   & 83.18 & 51.69 & 99.85 & 51.68 & 81.04 \\
          & Struc. + Func. & 100   & 100   & 100   & 100   & 100   & 100   \\
    \midrule
    \multirow{3}[2]{*}{Celer cBridge$\dagger$} 
          & Struc. & 72.08 & 74.74 & 74.74 & 74.74 & 72.08 & 73.85 \\
          & Func. & 89.87 & 89.87 & 74.51 & 74.51 & 84.76 & 81.34 \\
          & Struc. + Func. & 100   & 100   & 100   & 100   & 99.97 & 99.99 \\

    \bottomrule
    
    \multicolumn{8}{l}{\small $\dagger$ Among those three main bridges, one serves as the testing while the others as training. }\\
    
    \toprule
    \multirow{2}[2]{*}{\textbf{Testing}} & \multicolumn{1}{c|}{\multirow{2}[2]{*}{\textbf{Features}}} 
    & \multicolumn{5}{c|}{\textbf{Evaluating Classifiers (Accuracy \%)}} &    \\
          &       & \textbf{LR} & \textbf{SVC} & \textbf{RF} & \textbf{AdaBoost} & \textbf{GN} & \textbf{Avg.}    \\
    \midrule
  
    \multirow{3}[2]{*}{Other 10 Bridges$\ddagger$}
        & Struc. & 75.00 & 79.00 & 79.00 & 79.00 & 75.00  & 77.67  \\
        & Func. & 42.50 & 47.00 & 51.00 & 48.00 & 39.50 & 43.75  \\        
        & Struc. + Func. & 99.00    & 99.00    & 100   & 100  & 87.50  & 97.58 \\
    \bottomrule
    
    \multicolumn{8}{l}{\small $\ddagger$ Randomly select ten bridges as testing while the main three bridges serve as training.} \\
    
    \end{tabular}%
    
  }
  \vskip -1ex
  \label{tab:M1_results}
\end{table}%

\subsection{Results of Withdrawal Matching (S2)}
\label{RQ2}

We evaluate the effectiveness in matching withdrawal transactions with the following metrics: 
1) Matching Rate (MR). The percentage of associated results that match ground truth labels. The higher rate means a better traceability of bridge apps. 
2) Zero Hit Rate (ZHR). The percentage of results that no matching candidate transactions are found.
3) Wrong Hit Rate (WHR). The percentage of results that fail to locate the correct one among candidate transactions.
    


Table~\ref{tab:our_M2_result} demonstrates the experimental results of \myModel~in the withdrawal transaction matching. 
\myModel~successfully associates most of the transactions (more than 92\%) of different bridges among 21,105 transactions to Binance Smart Chain and 3,287 transactions to Polygon. In particular, nearly 98.99\% of the Multichain withdrawal transactions could be tracked.
As described in \S\ref{DateSource}, there are 13 bridges that provide cross-chain transaction tracking services. To illustrate \myModel's effectiveness outside these three bridges, we also collect cross-chain transaction pairs for the remaining 10 bridges and find that 96\% of the transactions are successfully matched.

Also, we construct the search space at several different time intervals and record the corresponding search space size and matching rate. 
The results are shown in Fig.~\ref{fig:parameter_analysis}.
The increasing time interval led to increasing search space and \myModel 's matching rate (reducing the zero hits mentioned in the previous paragraph). 
The matching rate increases faster when the time interval is small, but the matching rate stabilizes later. 
If the increase in time interval does not increase the matching rate, the inflection point is used to set the time interval. Note that the experimental results of \myModel~report the optimal time interval $\Delta$ in the search space constructor. 

\begin{table}[t]
\setlength{\abovecaptionskip}{0.2cm}
  \centering
  \caption{Results in withdrawal transaction matching.}
  \scalebox{0.62}{
  \hspace{-2ex}
    \begin{tabular}{c|ccc|c|ccc|c}
    \toprule
    \multicolumn{1}{c|}{\multirow{2}[2]{*}{\textbf{Bridges}}} &
      \multicolumn{4}{c|}{\textbf{Ethereum $\Rightarrow$ BSC}} &
      \multicolumn{4}{c}{\textbf{Ethereum $\Rightarrow$ Polygon}}
      \\
     &
      \multicolumn{1}{c}{\centering \textbf{MR}} &
      \multicolumn{1}{c}{\centering \textbf{ZHR}} &
      \multicolumn{1}{c}{\centering \textbf{WHR}} &
      \multicolumn{1}{|c|}{\textbf{\# All}} &
      \multicolumn{1}{c}{\centering \textbf{MR}} &
      \multicolumn{1}{c}{\centering \textbf{ZHR}} &
      \multicolumn{1}{c}{\centering \textbf{WHR}} & 
      \multicolumn{1}{|c}{\textbf{\# All}} 
      \\
    \midrule
    \rowcolor{gray!10} 
    Celer cBridge 
        & 95.35\% & 4.06\% & 0.59\% & 7,296
        & 95.65\% & 2.68\% & 1.67\%& 598 \\
    Multichain 
        & 98.99\% & 0.71\% & 0.30\% & 8,349
        & 94.08\% & 4.24\%& 1.68\%& 2,144 \\
    \rowcolor{gray!10} 
    Poly Network 
        & 92.55\% & 6.70\% & 0.75\% &  5,460  
        & 92.66\% & 6.79\%& 0.55\%&  545 \\

    \midrule
    \red{Other 10 Bridges}
        & 96.72\% & 1.24\% & 2.04\% & 3,965 
        & 96.24\% & 1.75\% & 2.01\% & 1,144 \\
    
        
    \bottomrule
    \end{tabular}%
    }
  \label{tab:our_M2_result}%
\end{table}%

\subsubsection{Compared with Methods for CeFi Bridges}
\label{RQ2:Existing}
Note that the existing methods are not supported for direct use on DeFi bridges. 
    \textbf{YKM Heuristic~\cite{Yousaf2019Tracing,Zhang2022CLTracer}} is designed for CeFi bridges to directly obtain the cross-chain deposit transactions and withdrawal transactions via off-chain centralized information. YKM Heuristic syncs all on-chain data and uses a time constraint to acquire all transactions as the deposit transaction search space, relying on centralized APIs to directly return the matching withdrawal transaction hashes (as described in Table~\ref{tab:existing_method} and motivation in \S\ref{DateSource}). 
    However, the centralized API of the bridge is against the decentralized nature of the DeFi bridge, thus the assumption in our problem is not API dependent (See \S\ref{Problem}).
    To meet the scenario of DeFi bridges, some adaptive changes are made for experiments: 1) filter the transactions of bridge contracts instead of syncing all data to speed up the experiments; 2) no internal APIs are provided in DeFi bridges, only heuristic rules. These changes have also been applied by the authors~\cite{Yousaf2019Tracing}. 
    \textbf{S1 $\times$ YKM Heuristic} is a variant of \myModel.
    Among transactions of bridge apps, the deposit and withdrawal transactions of bridge contracts are identified by our identification step. With the identified transactions, the YKM Heuristic is used instead of our matching step. 
The comparative results are shown in Table~\ref{tab:comparison}. 
Different methods are set with the same $\Delta$ parameter under the same bridge to ensure fairness.
The YKM Heuristic associates a minority of the cross-chain transaction pairs because it acquires a larger set of candidate transactions with similar time and amount but can not handle them. 
This result is consistent with the authors' findings~\cite{Yousaf2019Tracing} that the matching rates \textbf{\textit{``decreased significantly''}} without the usage of APIs.
The YKM Heuristic adding identification method also lacks the ability to track cross-chain transactions in DeFi bridges, because the message between bridge contracts remains unknown without log mining.
Under a similar time constraint, we add receiving address constraint and token amount to narrow the search space, supporting ERC20 tokens and eliminating reliance on CeFi bridge APIs. 
The results indicate the effectiveness and necessity of our log-based matching method in \myModel.

\begin{table}[t]
\renewcommand{\arraystretch}{1.4}
  \centering
  \caption{MR results of withdrawal transaction matching.}
  \scalebox{0.52}{
    \hspace{-2ex}
    \begin{tabular}{c|ccc|ccc}
    \toprule
    \multicolumn{1}{c|}{\multirow{2}[2]{*}{\textbf{\scalebox{1.2}{Methods}}}} &
      \multicolumn{3}{c|}{\textbf{\scalebox{1.2}{ETH $\Rightarrow$ BNB}}} &
      \multicolumn{3}{c}{\textbf{\scalebox{1.2}{ETH $\Rightarrow$ Polygon}}}
      \\
     &
      \multicolumn{1}{p{4em}}{\centering\textbf{\scalebox{1.2}{Celer} \scalebox{1.2}{cBridge}}} &
      \multicolumn{1}{l}{\centering\textbf{\scalebox{1.2}{Multichain}}} &
      \multicolumn{1}{p{5em}|}{\centering\textbf{\scalebox{1.2}{Poly} \scalebox{1.2}{Network}}} &
      \multicolumn{1}{p{4em}}{\centering\textbf{\scalebox{1.2}{Celer} \scalebox{1.2}{cBridge}}} &
      \multicolumn{1}{l}{\textbf{\scalebox{1.2}{Multichain}}} &
      \multicolumn{1}{p{5em}}{\centering\textbf{\scalebox{1.2}{Poly} \scalebox{1.2}{Network}}} 
      \\
    \midrule
    \rowcolor{gray!10} 
    \multicolumn{1}{c|}{\centering \scalebox{1.2}{YKM~\cite{Yousaf2019Tracing}}} &
      \scalebox{1.4}{32.50\%}&
      \scalebox{1.4}{0.70\%} &
      \scalebox{1.4}{2.30\%} &
      \scalebox{1.4}{61.04\%} &
      \scalebox{1.4}{7.56\%} &
      \scalebox{1.4}{1.28\%}
      \\
    \multicolumn{1}{p{7em}|}{\centering \scalebox{1.2}{\red{S1} $\times$ YKM~\cite{Yousaf2019Tracing}}} &
       \scalebox{1.4}{36.39\%}&
       \scalebox{1.4}{4.54\%}&
       \scalebox{1.4}{1.25\%}&
       \scalebox{1.4}{60.87\%}&
       \scalebox{1.4}{21.04\%}&
       \scalebox{1.4}{1.28\%}
      \\
      \rowcolor{gray!10} 
    \scalebox{1.2}{\myModel} &
      \scalebox{1.4}{95.35\%} &
      \scalebox{1.4}{98.99\%} &
      \scalebox{1.4}{92.55\%} &
      \scalebox{1.4}{95.65\%} &
      \scalebox{1.4}{94.08\%} &
      \scalebox{1.4}{92.66\%}
      \\
    \midrule
    \scalebox{1.2}{Time Interval ($\Delta$)} &
      \scalebox{1.4}{90} &
      \scalebox{1.4}{120} &
      \scalebox{1.4}{80} &
      \scalebox{1.4}{40} &
      \scalebox{1.4}{140} &
      \scalebox{1.4}{55}
      \\
    \bottomrule
    \end{tabular}%
  }
  \label{tab:comparison}%
\end{table}%

\begin{figure}[t]
\setlength{\abovecaptionskip}{0.2cm}
    \centering
    \hspace{-2.5ex}
	\subfigure[Celer cBridge]{%
        \centering
        \includegraphics[width=0.33\linewidth]{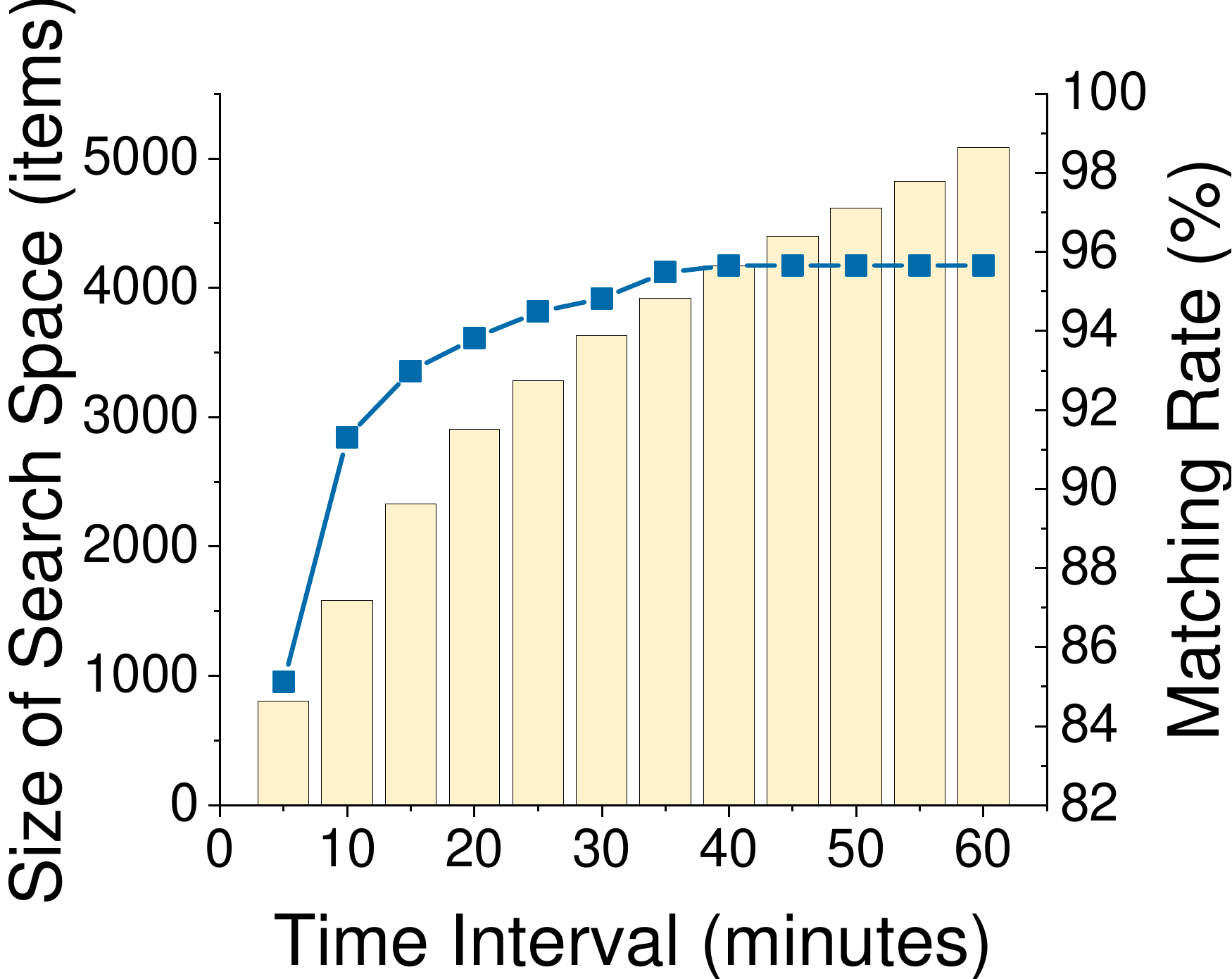}
	}
    \hspace{-0.6ex}
    \subfigure[Multichain]{%
        \centering
        \includegraphics[width=0.33\linewidth]{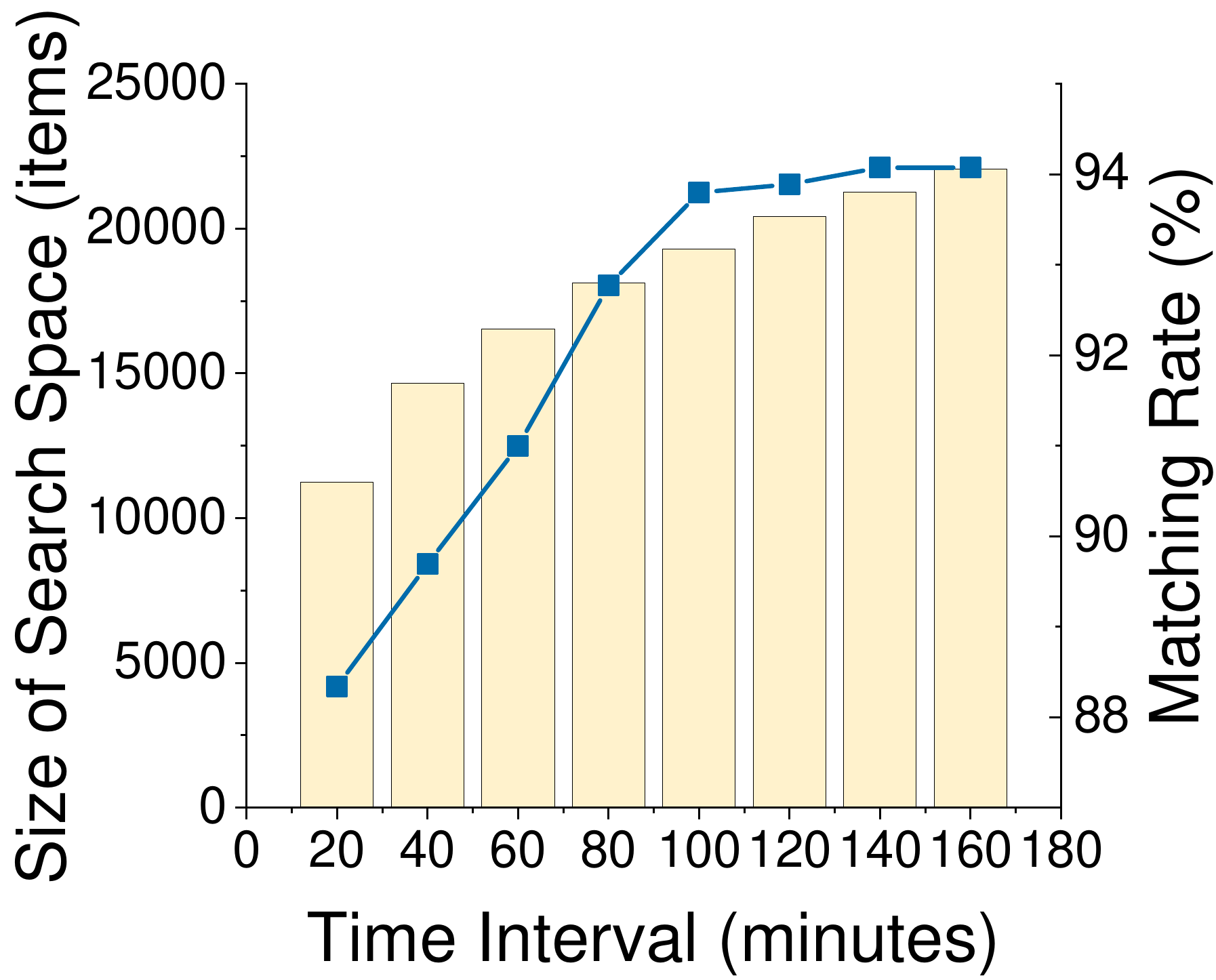}
	}    
    \hspace{-0.6ex}
	\subfigure[Poly Network]{%
        \centering
        \includegraphics[width=0.33\linewidth]{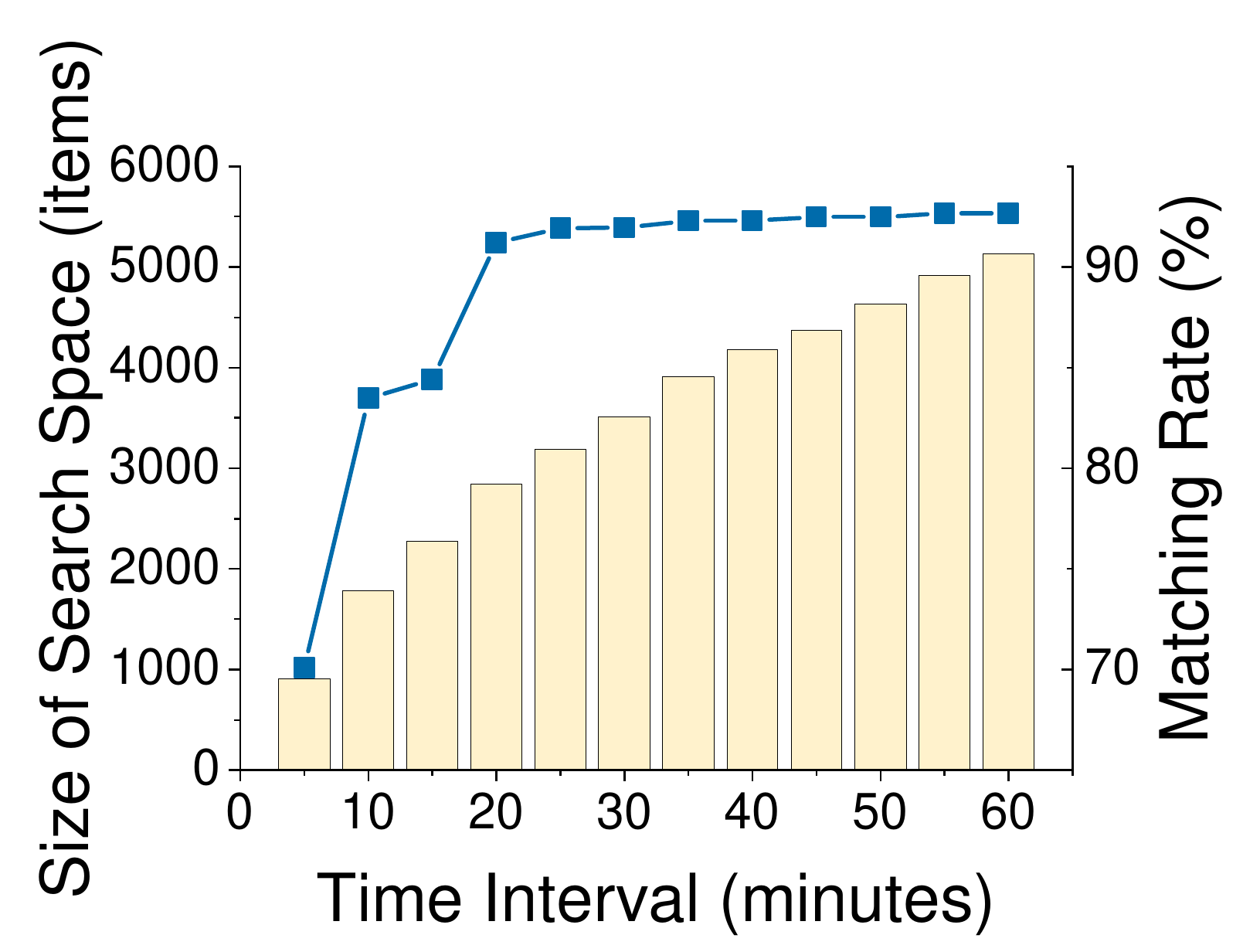}
	}

	\caption{The effect of time interval $\Delta$ on matching rate and  search space size (Ethereum $\Rightarrow$ Polygon). Inflection points of $\Delta$ in Celer cBridge, Poly Network, and Multichain are 40, 140, and 55, respectively.}
	\label{fig:parameter_analysis}
\end{figure}

\begin{figure}[t]
\setlength{\abovecaptionskip}{0.2cm}
    \centering
    \includegraphics[width=0.95\linewidth]{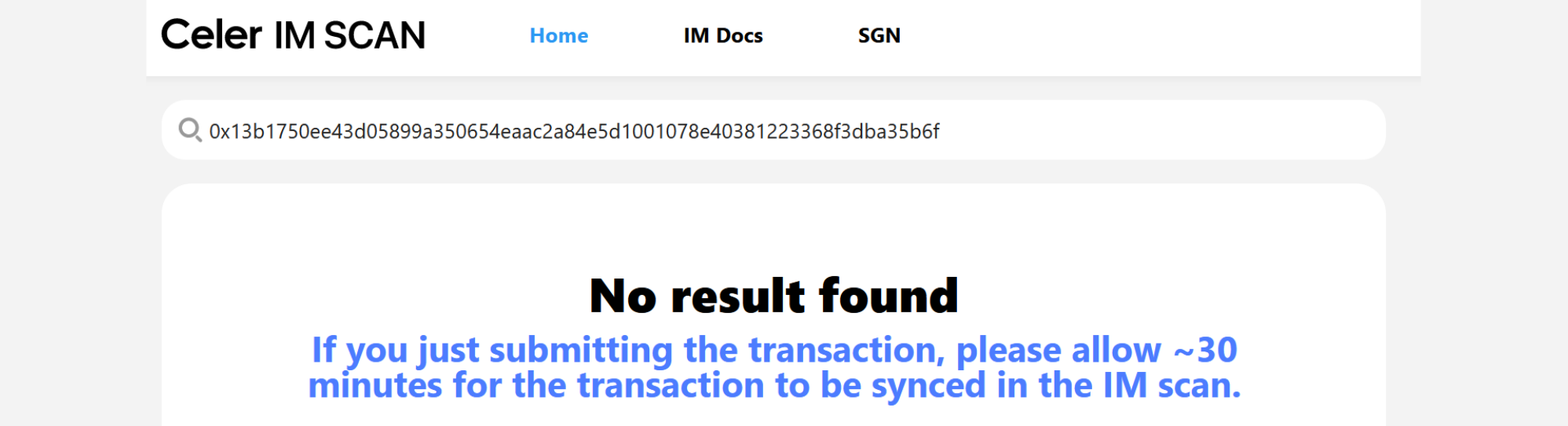}
    \caption{Fail to track and return \href{https://celerscan.com/tx/0x13b1750ee43d05899a350654eaac2a84e5d1001078e40381223368f3dba35b6f}{"No result found"}.} 
    \vskip -3ex
    \label{fig:celerscan_failed}
\end{figure}

\subsubsection{Compared with Explorers of DeFi Bridges}
\label{Case_study}

In our practical experiments, we observe the superiority of our \myModel~over bridge explorers. 
We conduct large-scale experiments on three bridges that offer cross-chain transaction tracking services to obtain the ground truth.
Particularly, we input the deposit transactions into the bridge browser and crawl the destination chain transactions returned.
However, we find that some deposit transactions yield results in our \myModel, while the bridge tracking service fails to return any results. 
For instance, on Jun-28-2021 11:37:11, a user on Ethereum initiated a cross-chain transaction to Celer cBridge, with the deposit transaction \href{https://etherscan.io/tx/0x13b1750ee43d05899a350654eaac2a84e5d1001078e40381223368f3dba35b6f}{\textit{0x13b1}}.
When we query this transaction hash in the bridge explorer, the result returned is \href{https://celerscan.com/tx/0x13b1750ee43d05899a350654eaac2a84e5d1001078e40381223368f3dba35b6f}{``No result found''}, as shown in Fig.~\ref{fig:celerscan_failed}.
By using \myModel, we find that this deposit transaction involves transferring 15 USDT tokens to Binance Smart Chain, and discover
a possible target withdrawal transaction \href{https://bscscan.com/tx/0x2fda9456be0b86302959aed985d85f321d627662ece8c03212b304995899d811}{\textit{0x2fda}} in which the user received 14.64255 BSC-USD on Jun-28-2021 11:40:42.
Upon manually inspecting the open-source documentation and implementation of Celer cBridge, we confirm that transaction \textit{0x2fda} is indeed a withdrawal transaction.
This case demonstrates that our \myModel~can achieve cross-chain transaction association, even in cases where bridge explorers do not provide support, further highlighting the utility of our \myModel.
Furthermore, we observe that this deposit transaction \href{https://etherscan.io/tx/0x13b1750ee43d05899a350654eaac2a84e5d1001078e40381223368f3dba35b6f}{\textit{0x13b1}} is executed by one of Celer cBridge's Ethereum contracts 
\href{https://etherscan.io/address/0x841ce48f9446c8e281d3f1444cb859b4a6d0738}{\textit{0x841c}}.
We also randomly select several transactions associated with this contract and none of them obtain the transaction association results from cBridge explorer. The reason may be due to an update of Celer cBridge to Version 2.0\footnote{\url{https://blog.celer.network/2021/11/29/cbridge-2-0-community-driven-upgrade-incoming/}}, rendering the old contract incompatible with the transaction association service.


To compare the efficiency of manual transaction association and \myModel, we track the earliest 50 transactions on Celer cBridge using both methods, with our \myModel~and the manual approach conducted by two experienced Ph.D. students familiar with cross-chain bridges.
Statistically, each transaction takes approximately 5-10 minutes of manual effort (using blockchain explorers), while our \myModel~(in deducting transaction fetching) takes just 0.84 seconds per transaction. This result not only shows that \myModel~can enhance cross-chain transaction association results (available in the open repository), but also improves the efficiency of cross-chain transaction association. 
%

\section{Empirical Analysis}
\label{Empirical}
In this section, we conduct an empirical analysis of cross-chain transaction pairs based on \myModel's output.

\subsection{Analysis of Identified Deposit Transactions}

\begin{figure}[t]
\setlength{\abovecaptionskip}{0.2cm}
    \centering
	\subfigure[Deposit transactions]{%
        \centering
        \includegraphics[width=0.45\linewidth]{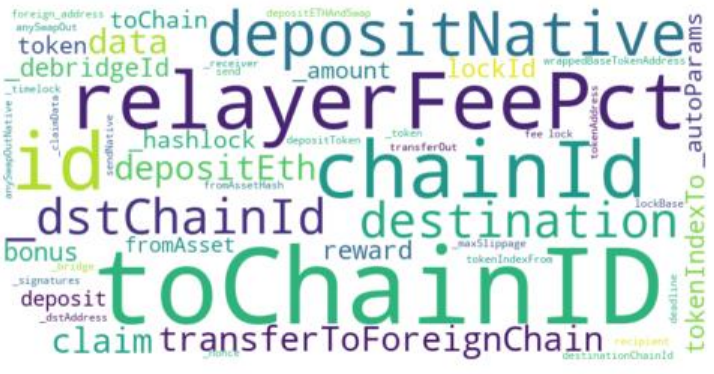}
	}
	\subfigure[Non-deposit transactions]{%
        \centering
        \includegraphics[width=0.45\linewidth]{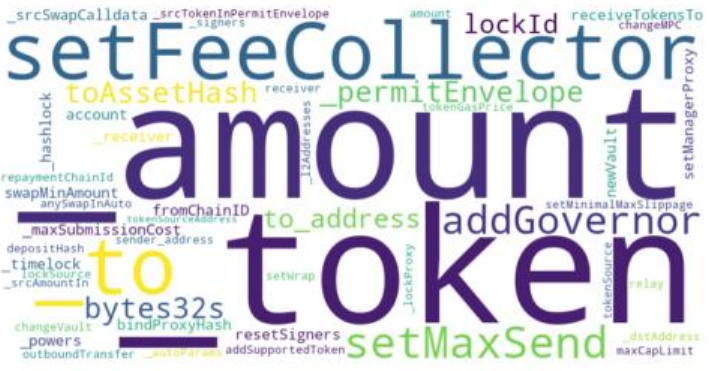}
	}
	\caption{Parameter cloud of transactions' input data.}
	\label{fig:input_wordcloud}
\end{figure}

\begin{figure}[t]
\setlength{\abovecaptionskip}{0.2cm}
    \centering
    \includegraphics[width=0.7\linewidth]{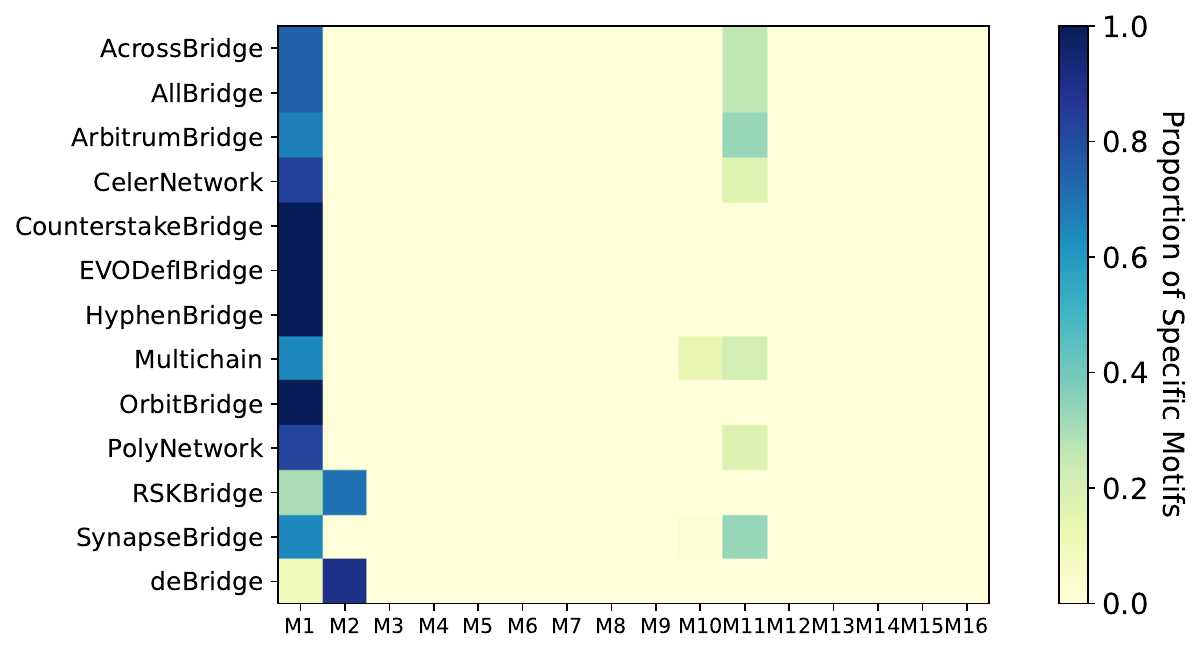}
    \caption{Distribution of the various motifs in deposit transactions of 13 DeFi bridges.}
    \label{fig:motif_analysis}
\end{figure}


\noindent\textbf{Functional Features.}
 We calculate the frequency of each element defined in \S\ref{word2vec} (function name or parameter name), and generate a word cloud map as shown in Fig.~\ref{fig:input_wordcloud}. It is easy to see that the
deposit transactions and non-deposit transactions are obviously different. The two most frequent elements of non-deposit transactions are related to the semantics of ``amount'' and ``token'', which frequently appear in token transfer transactions in normal DApps. Clearly, the most frequent elements of non-deposit transactions are related to the semantics of ``chain'' and ``relayer'', which are consistent with our pre-defined cross-chain metadata in \S\ref{Problem}.
However, extracting semantics directly from transaction function names and parameters provides strong yet limited information. Thus, \myModel~works well when deposit functions share similar names and align with metadata.
\begin{mdframed}[linewidth=0.7pt, linecolor=black, skipabove=6pt, skipbelow=6pt, backgroundcolor=gray!5]
\textbf{Finding 1.} The functional features of deposit transactions exhibit distinct characteristics, particularly in bridges that employ similar implementation, which enhances the accuracy of identification. 
\end{mdframed}

\noindent\textbf{Structural Features.}
We delve into the structural characteristics of deposit transactions of token-aware call graphs, by counting the numbers of transaction motifs (the motif types are shown in Fig.~\ref{fig:callgraph+motifs}(b). As illustrated in Fig.~\ref{fig:motif_analysis}, the subgraph structure of cross-chain transactions predominantly manifests as M$_1$, M$_2$, M$_{10}$, and M$_{11}$. M$_1$ represents the transaction pattern where accounts directly send native tokens to the bridge contract. M$_{2}$ reveals the process where accounts initiate a cross-chain request, after which the bridge directly transfers tokens to the recipient address. M$_{10}$ describes a more complex procedure where accounts use the bridge contract to convert native tokens into other tokens and complete the cross-chain deposit. In this process, the cross-chain routing contract exchanges specified tokens with the token contract and deposits the tokens into the cross-chain bridge token contract, thereby finalizing the deposit. 
However, call relationships may also be present in non-deposit transactions, reducing their significance. Nevertheless, its implicit nature makes it less susceptible to variations in function implementations.
\begin{mdframed}[linewidth=0.7pt, linecolor=black, skipabove=6pt, skipbelow=6pt, backgroundcolor=gray!5]
\textbf{Finding 2.} Structural features of token-aware call graphs in deposit transactions contain the main procedure of DeFi bridges in general, but not significantly enough.
\end{mdframed}

\begin{table}[t]
\renewcommand{\arraystretch}{1.3}
  \centering
  \caption{The number of matched transaction pairs that use the same/different account for cross-chain.}
  \hspace{-2ex}
  \scalebox{0.5}{
    \begin{tabular}{l|r|rrr}
    \toprule
    
    \textbf{\scalebox{1.4}{Bridge}} & \textbf{\scalebox{1.4}{\# $Pair_{same}$}} & \textbf{\scalebox{1.4}{\# $Pair_{diff}$}} & \textbf{\scalebox{1.4}{\# Unique}} & \textbf{\scalebox{1.4}{\# Cluster}}\\
    \midrule
    \rowcolor{gray!10} 
    \scalebox{1.4}{Allbridge Core} & \scalebox{1.4}{501 (95.79\%)} & \scalebox{1.4}{22 (4.21\%)} & \scalebox{1.4}{20 (90.91\%)} & \scalebox{1.4}{20 (90.91\%)}\\\scalebox{1.4}{Celer cBridge} & \scalebox{1.4}{7,884 (99.87\%)} & \scalebox{1.4}{10 (0.13\%)} & \scalebox{1.4}{2 (2\%)} & \scalebox{1.4}{2 (2\%)}\\
    \rowcolor{gray!10} 
    \scalebox{1.4}{Connext Bridge}  & \scalebox{1.4}{535 (96.40\%)} & \scalebox{1.4}{20 (3.60\%)} & \scalebox{1.4}{7 (35\%)} & \scalebox{1.4}{7 (35\%)}\\
    \scalebox{1.4}{Debridge} & \scalebox{1.4}{44 (91.67\%)} & \scalebox{1.4}{4 (8.33\%)} & \scalebox{1.4}{4 (100\%)} & \scalebox{1.4}{3 (75\%)}\\
    \rowcolor{gray!10} 
    \scalebox{1.4}{Multichain} & \scalebox{1.4}{10,253 (97.71\%)} & \scalebox{1.4}{240 (2.29\%)} & \scalebox{1.4}{177 (73.75\%)} & \scalebox{1.4}{150 (62.5\%)}\\
    \scalebox{1.4}{Poly Network} & \scalebox{1.4}{5,995 (99.83\%)} & \scalebox{1.4}{10 (0.17\%)} & \scalebox{1.4}{6 (60\%)} & \scalebox{1.4}{6 (60\%)}\\
    \rowcolor{gray!10} 
    \scalebox{1.4}{Router Protocol} & \scalebox{1.4}{17 (100.00\%)} & \scalebox{1.4}{0 (0.00\%)} & \scalebox{1.4}{-} & \scalebox{1.4}{-}\\
    \scalebox{1.4}{Stargate} & \scalebox{1.4}{1,210 (99.84\%)} & \scalebox{1.4}{2 (0.16\%)} & \scalebox{1.4}{2 (100\%)} & \scalebox{1.4}{2 (100\%)}\\
    \rowcolor{gray!10} 
    \scalebox{1.4}{Symbiosis} & \scalebox{1.4}{148 (98.01\%)} & \scalebox{1.4}{3 (1.99\%)} & \scalebox{1.4}{2 (66.67\%)} & \scalebox{1.4}{2 (66.67\%)}\\
    \scalebox{1.4}{Synapse Protocol} & \scalebox{1.4}{677 (98.26\%)} & \scalebox{1.4}{12 (1.74\%)} & \scalebox{1.4}{12 (100\%)} & \scalebox{1.4}{12 (100\%)}\\
    \rowcolor{gray!10} 
    \scalebox{1.4}{Transit Swap} & \scalebox{1.4}{471 (33.22\%)} & \scalebox{1.4}{947 (66.78\%)} & \scalebox{1.4}{688 (72.65\%)} & \scalebox{1.4}{630 (66.53\%)}  \\
    \scalebox{1.4}{Wormhole} & \scalebox{1.4}{476 (98.14\%)} & \scalebox{1.4}{9 (1.86\%)} & \scalebox{1.4}{5 (55.56\%)} & \scalebox{1.4}{5 (55.56\%)}\\
    \bottomrule
    \end{tabular}%
   }
  \label{tab:same_account}%
\end{table}%

\subsection{Analysis of Matched Transaction Pairs}
We explore the cross-chain transaction pairs output by \myModel, including issues about user anonymity protection, time efficiency, and economic benefits.

\noindent\textbf{User Anonymity Protection.}
We analyze the deposit account ($\mathbb{M}.sender$) and the withdrawal account ($\mathbb{M}.reciever$) of matched cross-chain transaction pairs, as shown in Table~\ref{tab:same_account}. Except for the Poly Network bridge, in the vast majority of cross-chain transactions, the sender of the deposit transaction and the receiver of the withdrawal transaction are both at the same address, and these transactions have unique addresses (`\# Unique' column). Especially for the Router Protocol bridge, the ratio of transactions that use the same address is 100\% (Assume that one entity uses only one account~\cite{Yousaf2019Tracing}.) This suggests that most cross-chain bridge users are not aware of privacy protection. Very few users use the new address as the recipient, and a deliberate effort to enhance anonymity could be involved in illegal financial activities such as money laundering (discussed in \S\ref{Methodology}). Further, ours performs further clustering analysis on cross-chain transaction pairs that employ different addresses (`\# Cluster' column). The clustering strategy groups accounts within the minimum connectivity graph into one class of accounts. We find that this portion of privacy-enhancing cross-chain transactions mainly comes from different users.

\begin{mdframed}[linewidth=0.7pt, linecolor=black, skipabove=6pt, skipbelow=6pt, backgroundcolor=gray!5]
\textbf{Finding 3.} Most cross-chain bridge users lack the awareness of anonymity privacy protection, using the withdrawal address of destination chains as same as the deposit account of source chains.
\end{mdframed}

\begin{figure}[t]
\setlength{\abovecaptionskip}{0.2cm}
    \centering
	\subfigure[Time interval of transaction pairs]{%
        \centering
        \includegraphics[width=0.48\linewidth]{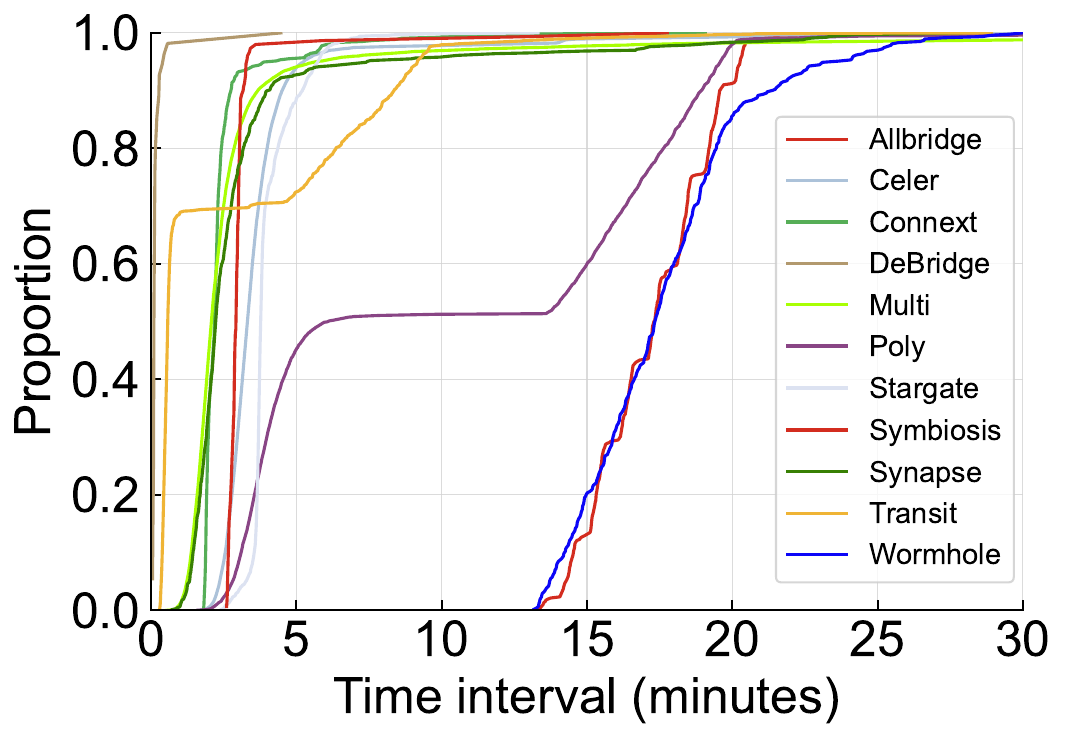}
	}
	\subfigure[Fee ratio of transaction pairs]{%
        \centering
        \includegraphics[width=0.48\linewidth]{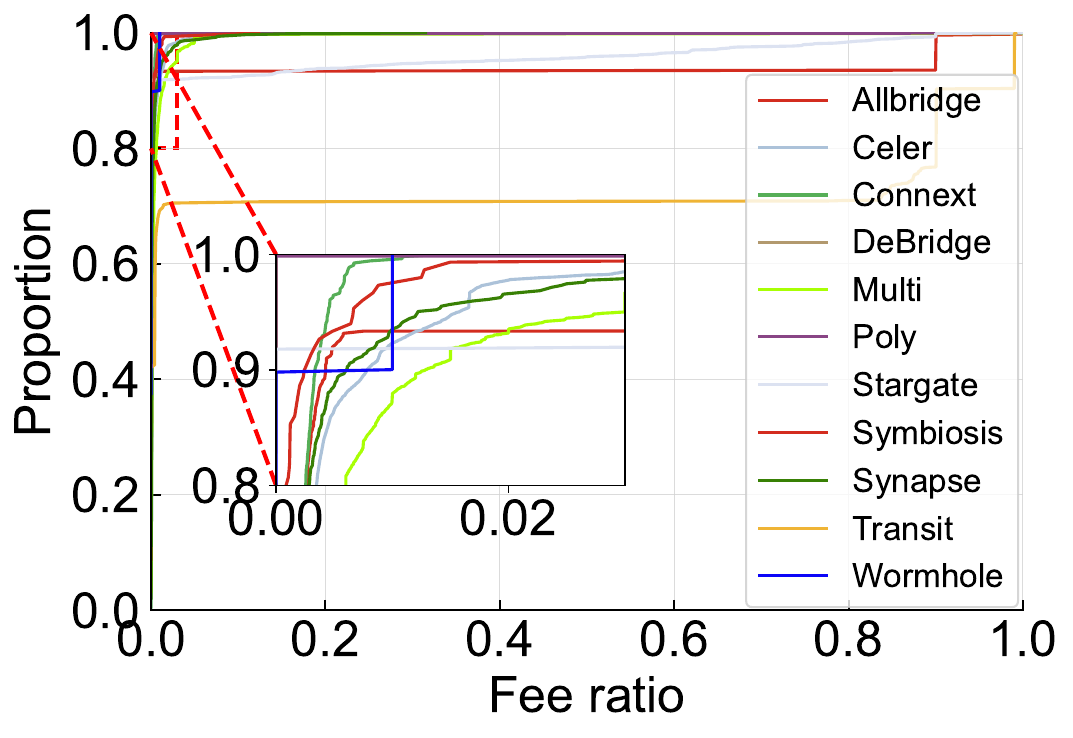}
	}
	\caption{Cumulative distribution of cross-chain transaction pairs among 13 DeFi bridges.}
    \vskip -3ex
	\label{fig:Proportion}
\end{figure}

\noindent\textbf{Time Efficiency.}
To explore the time efficiency of bridges across chains, we statistically analyze the deposit and withdrawal time intervals for each cross-chain transaction pair. According to Fig.~\ref{fig:Proportion}(a), we find that the vast majority of the destination chain's withdrawal transactions are able to be completed within 30 minutes, which validates the rationale for initially setting the business rule's time constraint $\tau$  to less than 30 minutes at our methodology level (Review details in {\sc Rule 2} of \S\ref{TargetDiscovery}). Further observation reveals that specific cross-chain bridges, such as DeBridge and Allbridge, perform better in terms of efficiency, with most of the transactions they support being completed quickly within 5 minutes. In contrast, other cross-chain bridges, such as Wormhole and Symbiosis, are mostly spread out in the 10 to 30-minute range. These reveal that there are some differences in the performance of different cross-chain bridges. The bridges encounter network congestion on the blockchain or additional validation time required when executing cross-chain transactions. 
In other cases, the user initiates more than one transaction of similar amounts simultaneously, which may result in wrong hits. Fortunately, these cases are unusual since most users make cross-chain transactions one at a time and on demand.

\begin{mdframed}[linewidth=0.7pt, linecolor=black, skipabove=6pt, skipbelow=6pt, backgroundcolor=gray!5]
\textbf{Finding 4.} The vast majority of withdrawal transactions can be completed in less than 30 minutes on destination chains,  but there may be instances where extra time intervals are required.  
\end{mdframed}

\noindent\textbf{Economic Benefits.}
To explore the economic benefits of cross-chain bridges, we statistically analyze the handling fee ratio for each cross-chain transaction pair, and obtain the results shown in Fig.~\ref{fig:Proportion}(b). The statistics show that 90\% cross-chain transaction pairs have a fee ratio of less than 1\%, while our amount constraint $\delta$ is set to 3\% (Review details in {\sc Rule 3} of \S\ref{TargetDiscovery}). However, we also observe that some of the transactions have fees that exceed the constraints, and speculate that these accounts may be willing to pay higher fees to speed up the process due to the urgent need for transaction speed. Note that we dynamically adjusted the detection range of the amount constraint $\delta$ to address the possible high fee share cases and improve the generality.
\begin{mdframed}[linewidth=0.7pt, linecolor=black, skipabove=6pt, skipbelow=6pt, backgroundcolor=gray!5]
\textbf{Finding 5.} Most cross-chain transaction pairs earn a fee ratio of less than 1\%. Users sometimes speed up cross-chain transactions by increasing transaction fees.
\end{mdframed}

\subsection{Potential Use in Anti-money Laundering}
\label{MoneyLaundering}

To verify the practical value of \myModel~in maintaining DeFi financial security, we use \myModel~to supplement cross-chain money laundering (ML) transactions for the security of decentralized finance.
Existing ML datasets, such as Elliptic (Bitcoin ML dataset)~\cite{elliptic2019Weber} and \textit{EthereumHeist} (Ethereum ML dataset)~\cite{XblockAML}, only identify ML transactions on single blockchain and do not include cross-chain transactions. 
First, we filter out bridge service contracts in the \textit{EthereumHeist} dataset~\cite{XblockAML} and find four security incidents, including: 
\textit{Akropolis Hack}
, \textit{Coinrail Hack}
, \textit{Fake\_Phishing5041}
, and \textit{Kucoin Hacker}
Then, for each security incident, we identify the cross-chain deposit ML transactions and obtain 5 deposit ML transactions for \textit{Akropolis Hack}, 5 for \textit{Coinrail Hack}, 1 for \textit{Fake\_Phishing5041}, and 15 for \textit{Kucoin Hacker}. We further utilize the matching module of \myModel~to discover related withdrawal transactions on the destination chain. 
Thus, with the use of our tool, security companies and regulatory authorities can supplement the cross-chain transfer paths of funds involved in these cases. 



\section{Threats to Validity}
\label{Threats}


\noindent\textbf{Internal validity.}
The bridge smart contract and deposit transaction labels for cross-chain transaction association may not be fully completed and reliable.
To mitigate the issue, two of the authors independently collected contracts and labeled the deposit transactions with the help of the third author to resolve a possible disagreement.
Also, bridges with totally disruptive function designs may be limited in the accuracy of learning-based identification. To mitigate this issue, structural features extracted from the token-aware call graph can make a complement, with accuracy stabilized at about 75\%.
The EOA-based CeFi bridge may exhibit unpredictable results without the bridge's APIs. The impact is limited because the percentage of CeFi bridges is quite small (as described in \S\ref{DateSource}).
The transaction tracking for inconsistent deposit-withdrawal transactions~\cite{Zhang2022Xscope} and many-to-many cross-chain relationships are not in the \myModel's scope. Those two issues belong to the research of attack detection~\cite{Zhou2022SoK} and coin mixing tracking~\cite{wu2021towards}, respectively.
Furthermore, our research does not encompass non-fungible token (NFT) bridges, privacy-enhancing bridges, or non-EVM-compatible bridges, which warrant investigation in future studies.

\noindent\textbf{External validity.}
\myModel's effectiveness largely depends on the availability of open-source contract codes. Bridges lacking these codes may pose challenges for our method. Among the contract-based DeFi bridges, 79.35\% of bridges disclose their contract addresses or codes (as investigated in \S\ref{DateSource}). Meanwhile, developers are usually willing to open their code so that whitehats can look for vulnerabilities. 

\section{Related work}
\label{Related}

\noindent\textbf{Tracing cross-chain transactions.} Bridging apps enable transfers of crypto assets across chains, making the routes of funds more difficult to trace. Tracing cross-chain transactions therefore becomes important for needs such as anti-money laundering and increased transaction transparency.
Yousaf \textit{et al.}~\cite{Yousaf2019Tracing} is the first to utilize the internal API and heuristic matching rules to achieve bridge traceability. 
Zhang \textit{et al.}~\cite{Zhang2022CLTracer} devised CLTracer, a framework grounded in address relationships for cross-chain transaction clustering. 
Existing methods cannot be directly applied to DeFi bridge traceability without centralized ledgers or APIs.

\noindent\textbf{Blockchain transaction analysis.}
Smart contracts execute their functions by receiving transactions that include different actions and semantics.
Currently, there is a lot of research focused on the transactions and behaviors of smart contracts~\cite{299645,9426396}. 
For example, 
Wang \textit{et al.}~\cite{9996367} used a sequence-based method to detect attack transactions in DeFi.
Wu \textit{et al.}~\cite{Wu2023MoTs} proposed a transaction semantic graph for transaction identification.
Zhou \textit{et al.}~\cite{DAppHunter2023} identified inconsistent behaviors of DApps by contrasting the behaviors from front-end, wallet, and smart contract.

\section{Conclusion}
\label{Conclusion}

In this paper, we propose \myModel, the first automatic transaction association tool designed for contract-based DeFi cross-chain bridges. 
\myModel~extracts features from transaction data and traces to accurately identify the deposit transactions and mine the executed logs of bridge contracts to match the correct withdrawal transactions. 
We build the first dataset with 24,392 cross-chain transaction pairs from bridge apps and conduct experiments. \myModel~successfully identifies 100\% deposit transactions, associates 95.81\% withdrawal transactions, and is superior to the bridge official website in practical use. 
Furthermore, \myModel~is able to associate additional cross-chain transactions than bridge official explorers, which demonstrates the superiority of \myModel. Based on the association results, we explore the issues about user privacy protection, time efficiency, and economic benefits of the cross-chain bridges, and we find that DeFi bridges are abused in crypto money laundering.
In summary, \myModel~enables bridge developers, regulators, and users to enhance bridge traceability in the multi-chain DeFi ecosystem.


\bibliographystyle{plain}
\bibliography{main}

\end{document}